\pdfoutput=1
\newif\ifjfm
\ifdefined\jfmversion \jfmtrue \fi

\ifjfm
  \documentclass[lineno]{JFM-FLM_Au}
  \usepackage{amsmath,bm}
  \usepackage{xcolor}
  \AtBeginDocument{\ifdefined\patchBothAmsMathEnvironmentsForLineno
    \LineNumberstrue\patchBothAmsMathEnvironmentsForLineno{equation}\fi}
\else
  \documentclass[11pt]{article}
  \usepackage[a4paper,left=2.6cm,right=2.6cm,top=2.6cm,bottom=2.6cm]{geometry}
  \usepackage{graphicx,amsmath,amssymb,bm}
  \usepackage[round,authoryear]{natbib}
  \usepackage[font=small,labelfont=bf]{caption}
  \usepackage{url}
  \usepackage{booktabs}
  \usepackage{times}
  \usepackage{xcolor}
  \definecolor{linkblue}{rgb}{0.10,0.25,0.55}
  \usepackage[colorlinks=true,allcolors=linkblue]{hyperref}
  \usepackage{titlesec}
  \titleformat{\section}{\normalfont\bfseries}{\thesection.}{0.6em}{}
  \titleformat{\subsection}{\normalfont\itshape}{\thesubsection.}{0.6em}{}
  \numberwithin{equation}{section}   %
  
\fi
\usepackage{tikz}
\usetikzlibrary{arrows.meta,positioning,calc,decorations.pathmorphing}

\newcommand{\backsec}[2]{%
  \ifjfm\begin{bmhead}[#1]#2\end{bmhead}\else\section*{#1}#2\fi}

\newcommand{\D}{\mathrm{d}}
\newcommand{\I}{\mathrm{i}}
\newcommand{\E}{\mathrm{e}}
\newcommand{\Ad}{\mathcal{A}}
\newcommand{\Zt}{Z_t}
\newcommand{\kb}{\bar{k}}

\newcommand{\Wup}{\mathcal{W}}          %
\makeatletter
\def\Xint#1{\mathchoice
  {\XXint\displaystyle\textstyle{#1}}%
  {\XXint\textstyle\scriptstyle{#1}}%
  {\XXint\scriptstyle\scriptscriptstyle{#1}}%
  {\XXint\scriptscriptstyle\scriptscriptstyle{#1}}%
  \!\int}
\def\XXint#1#2#3{{\setbox0=\hbox{$#1{#2#3}{\int}$}
    \vcenter{\hbox{$#2#3$}}\kern-.5\wd0}}
\makeatother
\newcommand{\pvint}{\Xint-}             %

\ifjfm
  \title{Compressible unsteady aerodynamics of finite-chord porous aerofoils}
  \author{Seongkyu Lee\aff{1}}
  \affiliation{\aff{1}Department of Mechanical and Aerospace Engineering,
    University of California, Davis, CA 95616, USA}
  \corresau{Seongkyu Lee, \href{mailto:skulee@ucdavis.edu}{skulee@ucdavis.edu}}
  \righttitle{Compressible unsteady aerodynamics of porous aerofoils}
  \lefttitle{S. Lee}
\else
  \title{\bfseries Compressible unsteady aerodynamics of\\
  finite-chord porous aerofoils}
  \author{Seongkyu Lee\\[4pt]
    {\small Department of Mechanical and Aerospace Engineering,}\\
    {\small University of California, Davis, CA 95616, USA}\\
    {\small \href{mailto:skulee@ucdavis.edu}{skulee@ucdavis.edu}}}
  \date{}
\fi

\begin{document}
\maketitle

\begin{abstract}
\noindent
A unified linear theory is developed for the unsteady loading of finite-chord
aerofoils, rigid or porous, in compressible subsonic flow. The formulation combines Possio's
integral operator with a convective permeable boundary condition, allowing
chordwise-varying admittance while retaining the wake and unsteady Kutta
condition. The loading exponents at aerofoil edges and admittance
discontinuities keep their incompressible form in terms of the local permeability
parameter, so weighted-Jacobi collocation carries over. The solution is verified against published incompressible results and an
independent compressible formulation. Gust and heave responses
and their indicial counterparts reveal that permeability controls the
sensitivity of unsteady loading to compressibility. As permeability increases,
the material impedance rather than the surrounding flow sets the pressure jump,
but only gradually: at Mach number 0.7 the gust load on a weakly permeable
surface changes by a third to a half, and closed-form steady and high-frequency
limits for resistive surfaces show that it vanishes only when the
permeability parameter greatly exceeds the Mach number. The
porous-to-impermeable load ratio, by which a treatment is judged, cannot be obtained by
combining incompressible porous and compressible rigid theories, which
misjudge it by up to a factor of two. Incompressible theory overestimates
this ratio for an acoustically compact chord and underestimates it for a
non-compact one, by factors of 1.4--4.3 at Mach numbers 0.5--0.7 and reduced
frequencies 5--20, overstating the load reduction. Since compactness depends on
Mach number times reduced frequency, the error persists at low speed, exceeding
10\,\% at Mach number 0.05 for reduced frequencies above 10.
\end{abstract}
\ifjfm
\fi

\section{Introduction}
\label{sec:intro}

An aerofoil in a non-uniform or unsteady stream carries an unsteady load, and
the classical theory of that load is the foundation of aeroelasticity, of rotor
and propeller load prediction, and of every engineering model of
turbulence-interaction noise \citep{bisplinghoff1955,glegg2017,moreau2019}. It
comprises \citet{theodorsen1935} for a section oscillating in a uniform stream,
\citet{sears1941} and \citet{vonkarman1938} for a section meeting a convected
gust, \citet{wagner1925} and \citet{kussner1936} for the corresponding indicial
responses, and \citet{possio1938} for the compressible kernel. All of it assumes
an impermeable surface.

Permeable surfaces are of interest because the loading they carry at an edge is
weaker. \citet{ffowcs1970} showed that the efficiency with which a half-plane
scatters near-field pressure into sound is controlled by the inverse-square-root
singularity of the loading at a rigid edge; softening that singularity is the
mechanism behind every porous edge treatment, and porous leading and trailing
edges have been measured to reduce broadband noise appreciably
\citep{geyer2011,chaitanya2020,bowen2022}. Porosity is one of a family of passive edge treatments, alongside serrations,
brushes, and compliant and graded surfaces, but the family does not act through
a single mechanism. A large part of the benefit of serrations comes from
destructive interference between sources distributed along an oblique edge, so
it is a property of the radiation and is retained in rotor predictions only when
the acoustic calculation keeps the edge geometry and the spanwise phase of the
loading \citep{kim2016,chaitanya2017,lyu2017,li2024}. Porous, poroelastic and
graded edges act instead on the edge singularity itself
\citep{jaworski2013,ayton2016}, so their benefit is carried by the unsteady load,
and the reduction predicted in application depends on how the treatment is
represented in the loading model rather than in the acoustic analogy that
follows it \citep{lee2021tbl}. The same singularity that sets the scattering,
moreover, is what generates leading-edge suction, so for this class of
treatment the aerodynamic price is fixed by the same quantity as the benefit.
Both statements are statements about \emph{loading}, and both therefore belong
to unsteady aerodynamics rather than to acoustics. That is the subject of this
paper: what a permeable surface does to the unsteady load on a finite chord in
compressible flow, before any acoustic analogy is applied.

Analytical treatments of permeable surfaces divide by geometry.
\citet{jaworski2013} solved the semi-infinite problem by Wiener--Hopf, with the
static porosity condition of \citet{howe1998}, in which the flow through the
surface is proportional to the pressure jump across it. They showed that, in
certain parametric limits, the far-field acoustic power of a porous edge scales
with the sixth power of the flow speed rather than the fifth power found for a
rigid edge, a result that motivated much of what followed. The finite-chord problem is different in kind:
the chord introduces a second edge, a shed wake and the Kutta condition, none of
which the half-plane carries. \citet{hajian2017} solved the steady finite-chord problem for a permeability
that varies along the chord, and \citet{baddoo2021} extended the analysis to
unsteady flow. They derived porous counterparts of the four classical unsteady
functions: the Theodorsen and Sears functions for harmonic oscillation and
sinusoidal gust encounter, and the Wagner and K\"ussner functions for the
corresponding indicial responses. The present analysis borrows its essential
device from \citet{baddoo2021}. At a permeable leading edge the load is still
singular, but more weakly than the inverse square root of a rigid edge. Its
exponent falls below $1/2$ by an amount set by the local permeability, and
returns to $1/2$ only where the surface is impermeable. A Chebyshev (Glauert) expansion
builds in the rigid exponent and cannot represent this behaviour, whereas a
family of Jacobi polynomials whose weights carry the correct exponents at the
two edges can. \citet{ayton2021} treated
chordwise-varying porosity by a Wiener--Hopf-based approximation over the same
frequency range.

Every one of those finite-chord treatments is incompressible, and the question
is whether that matters. For a rigid aerofoil the answer has long been known.
Consider a section of chord $2b$ in a stream of speed $U$, with sound speed
$c_0$ and Mach number $M = U/c_0$, responding at angular frequency $\omega$, so
that its reduced frequency is $\sigma = \omega b/U$ and the acoustic wavenumber
is $k_0 = \omega/c_0$. Incompressible gust theory requires the chord to be
acoustically compact, $k_0 b = M\sigma \ll 1$; beyond that the loading is no
longer in phase along the chord, and the compressible response departs from
Sears's function by more than a correction
\citep{graham1970,osborne1973,amiet1976,kemp1976}. The incompressible theories
are the limit $M \to 0$ at fixed $\sigma$ and so describe a compact chord at any
speed, since compactness is a condition on $M\sigma$, not on $M$. The surfaces
to which permeable edges are applied lie well outside that range, because they
move fast through disturbed flow. A rotor blade meets its own wake, a
neighbouring rotor's wake and the ambient turbulence once per revolution. At
$M = 0.5$--$0.7$, ingesting turbulence or passing through a tip vortex, it is
loaded at $\sigma = 5$--$20$, so $M\sigma$ lies between $2.5$ and $14$, and on
the advancing side of a high-speed rotor, where the local Mach number reaches
$0.6$--$0.9$, the impulsive part of the loading dominates what is radiated
\citep{jia2020}. A wing or fan blade at high subsonic speed that meets
atmospheric turbulence or the wakes of surfaces upstream is in the same
position: a gust whose wavelength equals the chord has $\sigma = \pi$, so at
$M = 0.5$--$0.7$ the product $M\sigma$ is already $1.6$--$2.2$, and shorter
gusts give larger values. Even a low-speed aerofoil is non-compact once $\sigma$
approaches $1/M$, about $20$ at $M = 0.05$, which is within the range that
turbulence-interaction and blade--vortex-interaction loading models must cover.
In each case the unsteady load is the input to aeroelastic and acoustic
predictions. For rotors and propellers it drives both the tonal and the
broadband parts of the signature, frequency-domain load models are what make the
prediction tractable at design time \citep{gill2026tonal}, and the spread in
predicted hovering-rotor noise that follows from gust-induced unsteady loading
is large enough that the loading model, not the acoustic propagation, sets the
uncertainty \citep{gill2026gust}. A treatment of such a surface therefore has
to be assessed in compressible flow, at finite chord, with the wake retained.

For a permeable aerofoil it is not known whether incompressible theory remains
adequate outside the compact range, and there is no reason to expect the rigid
criterion $M\sigma \ll 1$ to carry over.
The pressure jump across a rigid surface is set entirely by the surrounding
flow, which is where compressibility acts. Across a permeable surface it is set
partly by the surrounding flow and partly by the material impedance, which fixes
the pressure jump needed to drive fluid through the surface, and the balance
between the two shifts with permeability. How permeability controls the
sensitivity of the unsteady loading to compressibility, and how compressibility
changes the load of a porous aerofoil relative to that of an impermeable one,
which is how a treatment is judged, are the questions a compressible theory has
to answer.

At the compressible end, \citet{hales2024} formulated a semi-infinite permeable
edge in compressible flow with the grazing-flow conductivity of
\citet{howe1996}; being a half-plane theory it carries neither the finite chord
nor the wake. The combination that is missing (table~\ref{tab:capability}) is the subject of
this paper: a finite chord in compressible flow, with an admittance (the
reciprocal of the material impedance) that varies arbitrarily along the chord,
and with the wake and the Kutta condition.

\begin{table}
\centering\small
\begin{tabular}{@{}lcccc@{}}
\toprule
 & finite chord & compressible & wake, Kutta & arbitrary admittance\\
\midrule
\citet{jaworski2013}  & no  & partial & no  & no \\
\citet{hajian2017} (steady)   & yes & no  & n/a & yes \\
\citet{baddoo2021}    & yes & no  & yes & yes \\
\citet{ayton2021}     & yes & no  & yes & yes \\
\citet{hales2024}     & no  & yes & no  & no \\
present work          & yes & yes & yes & yes \\
\bottomrule
\end{tabular}
\caption{Analytical treatments of a permeable aerofoil surface. ``Partial''
records a theory formulated for compressible radiation from an incompressible
near field.}
\label{tab:capability}
\end{table}

This paper develops a linear theory for the unsteady loading of finite-chord
porous aerofoils in compressible subsonic flow. It combines Possio's integral
operator with a convective permeable boundary condition, allows the admittance
to vary along the chord, and retains the shed wake and the unsteady Kutta
condition. Three things change in passing from the incompressible formulation to
the compressible one, and one does not. The integral operator becomes Possio's
rather than Cauchy's, and its compressible part acts on the jump in velocity
potential itself and not only on the bound vorticity, so the compressible solver
must carry both where the incompressible one closes on the vorticity alone. The
permeable boundary condition becomes convective, so that the permeability
parameter (\S\ref{sec:bc}) carries the Mach number. And the reduced frequencies are stretched by
the Prandtl--Glauert transformation, so that the ratio of the chord to the
acoustic wavelength enters as well as its ratio to the hydrodynamic one. What does not change is the
local behaviour at the aerofoil edges and at admittance discontinuities. The
non-Cauchy part of Possio's operator is bounded, so it cannot alter the
exponents that a Riemann--Hilbert analysis assigns there, and the loading
exponents retain their incompressible functional form when expressed in terms of
the local permeability parameter (\S\ref{sec:edge}). This is what allows the
weighted-Jacobi collocation method of \citet{baddoo2021}, including their
treatment of interior admittance discontinuities, to be carried over at the cost
of a rigid calculation. The solution is verified
against published incompressible results and against an independent compressible
formulation in physical variables (\S\ref{sec:verify}). The formulation contains
the classical theories as limits: with $\lambda = 0$ it is Possio's problem, and
for $k_e \to 0$ it reduces to the functions of Theodorsen, Sears, Wagner and
K\"ussner; with $\lambda \neq 0$ and $k_e \to 0$ it is the incompressible porous
theory of \citet{baddoo2021}. Each limit is verified in \S\ref{sec:verify}.

The theory is then used to examine how permeability controls the sensitivity of
the unsteady loading to compressibility (\S\ref{sec:results}). Incompressible
and compressible responses are compared for the same material and reduced
frequency, both where the chord is acoustically compact and where it is not,
and at low as well as high Mach number, since compactness is set by $M\sigma$
rather than by $M$ (\S\ref{sec:lowmach}). The two routes by which the Mach
number reaches a permeable surface, the boundary condition and the integral
operator, are separated. The
shortcut of combining the existing incompressible porous and compressible rigid
theories is tested against the full solution, and the condition under which it
is adequate is derived. The harmonic gust and heave responses, the porous counterparts of the Sears and
Theodorsen functions, are computed together with their indicial counterparts,
the porous K\"ussner and Wagner functions, and the steady and high-frequency
limits are obtained in closed form. The aim is compressible loading predictions
for assessing porous aerofoils and for supplying aerodynamic inputs to
aeroelastic and aeroacoustic models.

The paper is organised as follows. Section~\ref{sec:formulation} formulates the
problem and derives the edge and junction exponents; \S\ref{sec:numerics}
describes the solution; \S\ref{sec:verify} verifies it against classical,
closed-form and published results; \S\ref{sec:results} compares the compressible
and incompressible responses; and \S\ref{sec:concl} draws conclusions.

\section{Formulation}
\label{sec:formulation}

\subsection{The problem}
\label{sec:problem}

\begin{figure}
\centering
\resizebox{0.96\textwidth}{!}{%
\begin{tikzpicture}[
  x=1cm,y=1cm,font=\footnotesize,
  gust/.style={draw=blue!62!black,line width=0.7pt,
               decorate,decoration={snake,amplitude=1.1mm,segment length=4.6mm}},
  ar/.style={-{Latex[length=2.1mm]},line width=0.6pt},
  dim/.style={{Latex[length=1.6mm]}-{Latex[length=1.6mm]},line width=0.6pt},
  lead/.style={draw=black!45,line width=0.6pt}
]
\def\LE{0}\def\SJ{2.05}\def\TE{6.6}\def\TH{0.075}

\fill[black!12] (\SJ,-\TH) rectangle (\TE,\TH);
\draw[line width=0.6pt] (\SJ,-\TH) rectangle (\TE,\TH);
\fill[blue!14] (\LE,-\TH) rectangle (\SJ,\TH);
\draw[line width=0.6pt] (\LE,-\TH) rectangle (\SJ,\TH);
\foreach \i in {1,...,11}{%
  \draw[blue!55!black,line width=0.6pt]
    ({\LE+\i*(\SJ-\LE)/12},-\TH) -- ({\LE+\i*(\SJ-\LE)/12},\TH);}
\fill[violet] (\SJ,0) circle (1.5pt);

\draw[dashed,line width=0.6pt] (\TE,0) -- (9.15,0);
\foreach \x in {6.95,7.4,7.85,8.3,8.75}{\draw[black!55,line width=0.6pt] (\x,-0.16) -- (\x,0.16);}
\node[anchor=west] at (7.25,0.58) {shed wake, $\Gamma\E^{\I\kb(\xi-1)}$};
\draw[lead] (7.60,0.44) -- (7.65,0.07);

\foreach \y in {2.05,2.50,2.95}{\draw[gust] (-3.75,\y) -- (-1.65,\y);}
\draw[ar,draw=blue!62!black] (-1.60,2.50) -- (-0.75,2.50);
\node[anchor=west,text=blue!62!black,align=left] at (-3.82,3.75)
  {convected gust, heave, pitch,\\[2pt] or a step: all enter as $\Wup(\xi)$};
\draw[ar] (-3.75,0.35) -- (-1.75,0.35);
\node[above] at (-2.75,0.39) {$U$, $M=U/c_0$};

\node[text=blue!62!black,anchor=south,align=center] at (0.05,1.02)
  {$\Pi\sim(1+\xi)^{-\beta}$\\[1pt] $\beta=\pi^{-1}\cot^{-1}\lambda(-1)$};
\draw[lead] (0.30,0.92) -- (0.14,0.14);
\node[text=violet,anchor=south,align=center] at (2.95,1.62)
  {$\Pi\sim|\xi-s_j|^{-\delta}$\\[1pt] $\delta=\beta_{\rm right}-\beta_{\rm left}$};
\draw[lead,draw=violet!60] (2.80,1.54) -- (\SJ,0.12);
\node[anchor=south,align=center] at (5.55,1.05)
  {$\Pi\sim(1-\xi)^{+\alpha}$\\[1pt] $\alpha=\pi^{-1}\cot^{-1}\lambda(+1)$};
\draw[lead] (5.95,0.98) -- (6.54,0.12);

\node[text=blue!62!black,anchor=north,align=center] at (1.00,-0.62)
  {permeable, $\lambda(\xi)\neq0$};
\node[anchor=north] at (4.30,-0.62) {rigid, $\lambda=0$};
\node[text=violet,anchor=north,font=\scriptsize] at (\SJ,-1.02) {junction $s_j$};

\draw[black!45,line width=0.6pt] (\LE,-\TH) -- (\LE,-2.10);
\draw[black!45,line width=0.6pt] (\SJ,-\TH) -- (\SJ,-0.98);
\draw[black!45,line width=0.6pt] (\TE,-\TH) -- (\TE,-2.10);
\draw[dim] (\LE,-1.50) -- (\SJ,-1.50);
\node[below] at ({(\LE+\SJ)/2},-1.48) {$\ell$};
\draw[dim] (\LE,-1.92) -- (\TE,-1.92);
\node[below] at ({(\LE+\TE)/2},-1.90) {$c=2b$};
\node[below,font=\scriptsize,text=black!60] at (\LE,-2.16) {$\xi=-1$};
\node[below,font=\scriptsize,text=black!60] at (\TE,-2.16) {$\xi=+1$};
\end{tikzpicture}
}%
\caption{The configuration and the three exponents. A thin section of chord
$c = 2b$ in a stream of Mach number $M$ carries a permeable region of length
$\ell$, across which the admittance enters through $\lambda(\xi)$ alone; here
the region begins at the leading edge and ends at a junction $s_j$, but the
formulation admits any distribution, including a permeable trailing edge. The
loading is singular at the leading edge and at the junction, bounded at the
trailing edge, and each exponent is an inverse cotangent of the local
$\lambda$ (\S\ref{sec:edge}). The wake and the unsteady Kutta condition make
the circulation part of the solution rather than a free parameter. A
semi-infinite formulation cannot represent this, and it is what makes the chord
enter the answer.}
\label{fig:geom}
\end{figure}
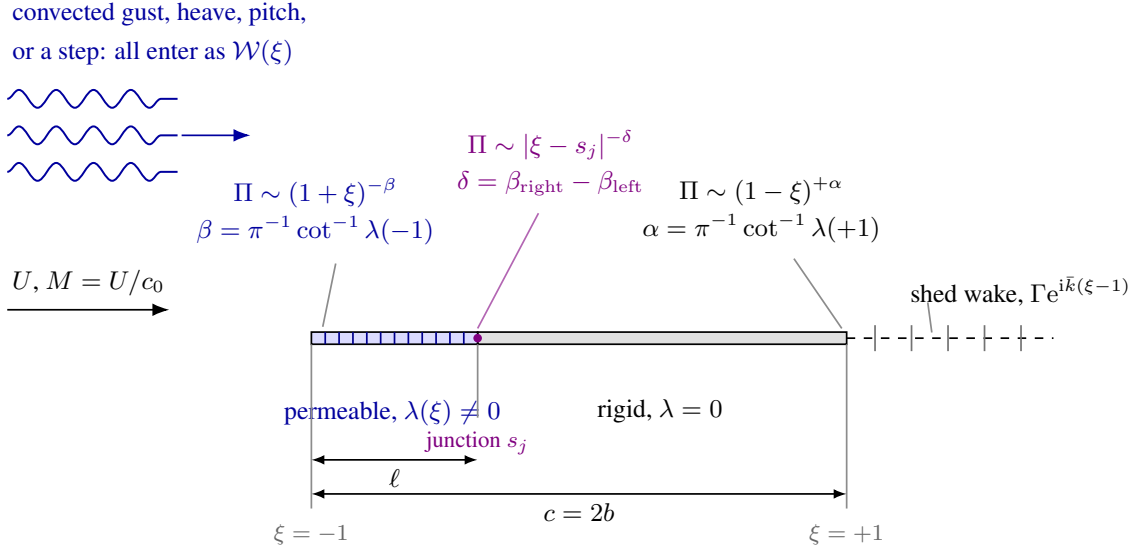

Figure~\ref{fig:geom} shows the configuration and the notation used below. Three features of the configuration
drive everything that follows. The permeable region has \emph{ends}, so the
problem carries interior discontinuities in boundary condition as well as the
two aerofoil edges, and \S\ref{sec:edge} shows that each carries its own
algebraic singularity with exponents that are not independent. The trailing
edge carries a shed wake and the unsteady Kutta condition, so the circulation
is fixed by the solution. And the admittance is a function of position, not a
constant: a manufactured insert has a definite extent, and the grazing flow
that sets the admittance of a perforate varies along it.

A thin aerofoil of chord $c = 2b$ lies along $-b \le x_1 \le b$, $x_2 = 0$, in a
uniform stream $U$ of an ideal fluid of density $\rho_0$ and sound speed $c_0$,
at Mach number $M = U/c_0$ and Prandtl--Glauert factor
$\beta_M = \sqrt{1-M^2}$. The surface is permeable, with a transfer admittance
that may vary along the chord. The disturbance is harmonic with the convention
$\E^{-\I\omega t}$; we write $k_0 = \omega/c_0$ for the acoustic wavenumber and
$k_1 = \omega/U$ for the hydrodynamic one. The canonical forcing is the frozen
transverse gust $w = w_0\exp[\I(k_1x_1 - \omega t)]$, but the formulation admits
any imposed upwash, and \S\ref{sec:forcing} lists the three that are used here.

The Prandtl--Glauert transformation (appendix~\ref{app:kernel}) maps the problem
to a Helmholtz problem of wavenumber $k_0/\beta_M$ on a plate of half-length
$b/\beta_M$. Normalising the chordwise coordinate on that stretched half-length,
$\xi \in [-1,1]$, the two reduced wavenumbers each pick up $\beta_M$ twice,
\begin{equation}
k_e = \frac{k_0 b}{\beta_M^2}, \qquad \kb = \frac{k_1 b}{\beta_M^2},
\qquad k_e = M\kb .
\label{eq:nondim}
\end{equation}
Both are dimensionless. $k_e$ is the phase, in radians, that a sound wave of the
transformed problem accumulates across its half-chord $b/\beta_M$, and so
measures acoustic compactness; $\kb$ is the corresponding phase of the convected
gust. Equivalently, the chord spans $k_e\beta_M^2/\pi$ acoustic wavelengths. The
classical reduced frequency is
$\sigma = \omega b/U = k_1 b = \beta_M^2\kb$, and results are reported in
$\sigma$ so that they may be read against the classical functions directly.

\subsection{The permeable boundary condition}
\label{sec:bc}

The surface is modelled as a homogenised sheet of vanishing thickness across
which the pressure jump $\Delta p$, upper surface minus lower, drives a seepage
velocity $v_s$, positive in the $x_2$ direction. Writing the normalised transfer
admittance $\Ad = 1/\Zt$ through
\begin{equation}
v_s = -\frac{\Ad\,\Delta p}{\rho_0 c_0},
\label{eq:vs}
\end{equation}
so that fluid passes from the side at higher pressure to the side at lower
pressure. Mass conservation across the sheet requires the normal velocity of
the fluid relative to the surface, the imposed upwash $w$ plus the disturbance
velocity $\partial\varphi/\partial x_2$, to equal the seepage velocity on both
faces, which gives
\begin{equation}
\frac{\partial\varphi}{\partial x_2}\bigg|_{x_2=0}
 = -w + v_s
 = -w + \frac{\Ad}{c_0}
   \left(\frac{\partial}{\partial t} + U\frac{\partial}{\partial x_1}\right)\!
   \big[\varphi\big],
\label{eq:bcconv}
\end{equation}
the second form following from the linearised momentum relation
$\Delta p = -\rho_0(\partial_t + U\partial_{x_1})[\varphi]$, with $[\varphi]$ the
jump in velocity potential. Written this way the condition is
\emph{convective}: the seepage is proportional to the material rate of change of
the potential jump seen by the mean flow, and not to the potential jump itself.
The convective operator sits in the momentum relation rather than in the
material law \eqref{eq:vs}, which is why much of the porous-media literature,
working at $U = 0$, sees only $\partial_{x_2}\varphi \propto [\varphi]$. The two
conditions coincide only when $U = 0$. The convective derivative comes from the
linearised Bernoulli relation, which carries it in incompressible flow as well,
so it is not removed by letting $M \to 0$ at fixed $U$.

With harmonic time dependence and lengths scaled on the stretched half-chord,
the material enters the integral equation of \S\ref{sec:sie} through a single
dimensionless group, the permeability parameter
\begin{equation}
\lambda = \frac{2\Ad M}{\beta_M},
\label{eq:lambda}
\end{equation}
which multiplies the pressure jump there. A rigid surface has $\lambda = 0$, a
fully transparent one $\lambda \to \infty$, and a chordwise distribution of
material is a function $\lambda(\xi)$. Appendix~\ref{app:bc}
gives the derivation and shows that $\lambda$, being a ratio of the seepage
coefficient to the Cauchy coefficient, does not depend on how the unknown is
normalised. It also shows how the condition relates to the static condition of
\citet{jaworski2013} and \citet{ayton2021}: holding the porosity parameter $\mu
= -\I k_0\Ad$ fixed as $M \to 0$ keeps $\lambda = 2\I\mu/(\beta_Mk_1)$ finite,
but the static condition is recovered only when, in addition, the convective
derivative is dropped.

\paragraph{Relation to the porosity parameter of \citet{baddoo2021}.} Their
$\psi = 4/(\Phi + 2\I k\rho_e)$, with $\Phi$ the dimensionless resistance,
$\rho_e$ the effective density of the pore fluid and $k = \omega b/U$ their
reduced frequency, equal to $\sigma$, is the same quantity as
$\lambda$, written in the time convention $\E^{+\I kt}$ and for a specific
material law. The identification is not an analogy: it is an equality of the
coefficient multiplying the loading in the two governing equations, it is
confirmed numerically in \S\ref{sec:verify} across the full unsteady range, and
it is why their exponent formula and ours are the same function. In the present
convention the same material is the conjugate,
$\psi \mapsto 4/(\Phi - 2\I k\rho_e)$.

\subsection{Reduction to a singular integral equation}
\label{sec:sie}

Representing the aerofoil and its wake by a dipole sheet whose strength is the
jump in potential, taking the pressure jump $\Delta p(\xi)$ as the unknown, and
enforcing \eqref{eq:bcconv} on the mean surface yields,
after the Prandtl--Glauert transformation,
\begin{equation}
\lambda(\xi)\,\Pi(\xi) + \frac{1}{\pi}\pvint_{-1}^{1}\frac{\Pi(s)}{\xi-s}\,\D s
\;+\; \int_{-1}^{1} \mathcal{K}(\xi-s)\,\Pi(s)\,\D s \;=\; \Wup(\xi),
\label{eq:sie}
\end{equation}
where $\Pi$ is the pressure jump, non-dimensionalised so that the Cauchy
coefficient is $1/\pi$, and $\mathcal{K}$ is a bounded remainder. The bar on the
integral sign denotes the Cauchy principal value,
\[
\pvint_{-1}^{1}\frac{\Pi(s)}{\xi-s}\,\D s
\;\equiv\; \lim_{\epsilon\to0^{+}}\left(
\int_{-1}^{\xi-\epsilon} + \int_{\xi+\epsilon}^{1}\right)
\frac{\Pi(s)}{\xi-s}\,\D s ,
\]
which exists because $\Pi$ is H\"older-continuous on any closed subinterval of
$(-1,1)$ that contains no admittance junction, and is integrable across the
junctions (\S\ref{sec:junction}). The operator is the Possio kernel \citep{possio1938}:
appendix~\ref{app:kernel} derives \eqref{eq:sie} from the convected wave
equation and shows that its singular part is exactly the Glauert Cauchy kernel,
while the remainder consists of the regularised Hankel function
$H_1^{(1)}(z) + 2\I/(\pi z)$, which is bounded, together with terms whose
kernels are at worst logarithmically singular. Each therefore maps a
H\"older-continuous $\Pi$ to a bounded function, which is all the analysis below
requires; $\mathcal{K}$ is a bounded \emph{operator} rather than a bounded
convolution kernel. It includes the wake, which enters through the circulation;
at the trailing edge the chord and wake contributions each carry a logarithm,
and appendix~\ref{app:kernel} shows that the two cancel, so that $\mathcal{K}$
stays bounded there too.

Two auxiliary conditions close the problem: Kelvin's theorem, relating the bound
circulation to the shed vorticity, and the unsteady Kutta condition, which
requires the loading to be bounded at the trailing edge. On a rigid aft section
the latter is $\Pi \sim (1-\xi)^{1/2}$; on a permeable one it is
$\Pi\sim(1-\xi)^{+\alpha}$ with $\alpha$ given by \eqref{eq:alpha} below, which
is weaker but still bounded.

\paragraph{Provenance.} The equation is assembled from two independent
ingredients and is not inherited whole from any previous paper. The integral
operator is the compressible thin-aerofoil operator of \citet{possio1938},
unchanged. The term $\lambda(\xi)\Pi(\xi)$ is what the permeable boundary
condition \eqref{eq:bcconv} contributes, and is derived in
appendix~\ref{app:bc}. The relation to the incompressible equation of
\citet{baddoo2021} is a change of dependent variable together with the loss of the Hankel
remainder: they carry the bound vorticity $\gamma$, so their algebraic term
appears as $\gamma + \I k\!\int_{-1}^{\xi}\gamma$, a Volterra contribution which
is exactly their pressure jump, and writing the equation in $\Pi$ from the
outset absorbs it. Setting $k_e = 0$ removes both Hankel terms, and what remains
of $\mathcal{K}$ is precisely the bounded term generated by the change of
dependent variable; the present equation and theirs are therefore the same
equation written in different variables. Both are dominant-singular
equations with a variable coefficient, which is why they share edge exponents.

\subsection{The forcing}
\label{sec:forcing}

$\Wup$ is the forcing of \eqref{eq:sie} and the only place the incident flow
enters. It is the physical upwash $w(\xi)$, normalised on its amplitude, with
the sign it carries in \eqref{eq:bcconv}. Equation \eqref{eq:sie} is written for
the transformed potential of appendix~\ref{app:kernel},
$\varphi = \Psi\,\E^{-\I Mk_e\xi}$, so $w$ enters it multiplied by
$\E^{\I Mk_e\xi}$:
\begin{equation}
\Wup(\xi) = -\,w(\xi)\,\E^{\I Mk_e\xi},
\qquad
w(\xi) =
\begin{cases}
\E^{\I\sigma(\xi+1)} & \text{convected gust (Sears),}\\[2pt]
1 & \text{heave (Theodorsen),}\\[2pt]
1 - \I\sigma(\xi - \xi_p) & \text{pitch about $\xi = \xi_p$.}
\end{cases}
\label{eq:forcing}
\end{equation}
The heave upwash is normalised on the plunge velocity and the pitch upwash on
$U$ times the pitch amplitude, whose two terms are the incidence and the pitch
rate. The sign is the statement that the surface must cancel the imposed upwash,
up to whatever seepage $\lambda$ permits. For the gust the two exponentials
combine, $\sigma + Mk_e = \kb$, into $-\E^{\I(\kb\xi + \sigma)}$: in the
transformed variables the gust convects at $\kb$, the wavenumber at which the
wake convects in every case. In the heave and pitch problems too, $\kb$ enters through the wake and through
the loading $\Pi = -\partial m/\partial\xi + \I\kb m$ of \eqref{eq:PiFromM}, not
through the forcing. The gust phase is referred to the \emph{leading edge},
$\xi = -1$, and not to the midchord: this costs nothing in the frequency domain
but is essential when the transfer function is inverted to the time domain,
since a midchord reference makes the response begin before the gust arrives.
The indicial responses are not a fourth forcing: they are the heave and gust
responses carried into the time domain (\S\ref{sec:indicial}). In every case the
Kutta and Kelvin closures of \S\ref{sec:linsys} are unchanged.

The same factor appears, inverted, in the output. $\Pi$ is the transformed
loading; the physical pressure jump, upper surface minus lower, is
$\Pi(\xi)\,\E^{-\I Mk_e\xi}$, and the lift is
\begin{equation}
L = -\int_{-1}^{1}\Pi(\xi)\,\E^{-\I Mk_e\xi}\,\D\xi ,
\label{eq:lift}
\end{equation}
the sign because a lifting section carries the lower pressure on its upper
surface. The steady lift of appendix~\ref{app:steady}, $C_L = -2\int\Pi\,\D\xi$,
is the same statement. With the normalisation of appendix~\ref{app:bc},
$\Delta p = (2/\beta_M)\rho_0Uw_0\,\Pi\,\E^{-\I Mk_e\xi}$, the lift per unit
span is $(2/\beta_M)\rho_0Uw_0b\,L$, so that on the basis used in
\S\ref{sec:results}, the incompressible quasi-steady lift $2\pi\rho_0Uw_0b$ of
the rigid plate, the load is $L/(\pi\beta_M)$.
The factor has unit modulus, so it leaves the magnitude of the loading, and with
it every exponent of \S\ref{sec:edge}, unchanged; it does not enter the
permeable term either, since \eqref{eq:vs} relates the seepage to the pressure
jump at the same point and the factor multiplies both. It does change the lift
once $Mk_e$ is of order one, and a check that it has been carried consistently
is the high-frequency limit of the heave response, which must approach the
acoustic (piston) value, $2\beta_M/(\pi M)$ times the compressible quasi-steady
lift (table~\ref{tab:verify}).

An oblique gust, with spanwise wavenumber $k_3$, does not require a new
formulation either: the factor $\E^{\I k_3x_3}$ is a parameter of the
two-dimensional problem and enters only by replacing the acoustic wavenumber
with $\tilde{k}_e = (b/\beta_M)\sqrt{k_0^2/\beta_M^2 - k_3^2}$, the branch chosen
so that $\operatorname{Im}\tilde{k}_e \ge 0$. This matters for the acoustic use
of the theory and is recorded here for completeness; all results below are for
$k_3 = 0$.

\subsection{Edge conditions}
\label{sec:edge}

Equation~\eqref{eq:sie} is a dominant-singular equation with a variable
coefficient. Its solution is not smooth, and the local exponents are set by the
coefficient.

Near an edge the bounded part of the kernel is subdominant and \eqref{eq:sie}
reduces to the dominant equation on a half-line with constant coefficient. This
is a Riemann--Hilbert problem \citep{muskhelishvili1953,gakhov1966}. Writing the
sectionally analytic function
$\Omega(z) = (2\pi\I)^{-1}\int\Pi(s)/(s-z)\,\D s$ and applying the Plemelj
formulae gives $\Omega^+ = G\,\Omega^- + \mathrm{(forcing)}$ with
\begin{equation}
G = \frac{\lambda + \I}{\lambda - \I}.
\label{eq:G}
\end{equation}
Setting $\lambda = \cot(\pi\beta)$ makes $G = \exp(2\pi\I\beta)$, and the
canonical solution then behaves as $(1+\xi)^{-\beta}$ at the leading edge. Hence
\begin{equation}
\beta = \frac{1}{\pi}\cot^{-1}\lambda(-1)
      = \frac12 - \frac{\arctan\lambda(-1)}{\pi} .
\label{eq:beta}
\end{equation}
At the trailing edge the same construction with the admissible (bounded) branch
gives the exponent of the vanishing rather than of the singular behaviour,
\begin{equation}
\Pi \sim (1-\xi)^{+\alpha},
\qquad \alpha = \frac{1}{\pi}\cot^{-1}\lambda(1),
\label{eq:alpha}
\end{equation}
so that a rigid aft section, $\lambda(1) = 0$, recovers the classical
$(1-\xi)^{1/2}$, and a permeable one vanishes more slowly. Both exponents are
local: each is evaluated with the value of $\lambda$ at its own edge, and
neither depends on the rest of the distribution.

\begin{figure}
\centering
\includegraphics[width=\textwidth]{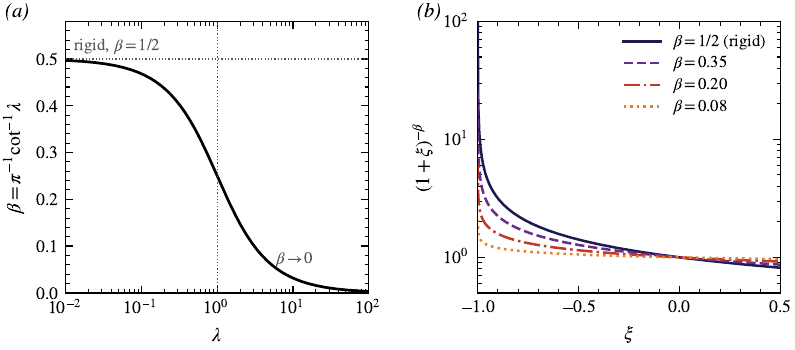}
\caption{($a$) The leading-edge exponent $\beta = \pi^{-1}\cot^{-1}\lambda$
of \eqref{eq:beta}. ($b$) The loading singularity $(1+\xi)^{-\beta}$ it
produces. A Glauert basis imposes $\beta = 1/2$, valid only for
$\lambda \ll 1$.}
\label{fig:edge}
\end{figure}

Two limits check \eqref{eq:beta}. As $\lambda \to 0$, $\beta \to 1/2$ and the
classical inverse-square-root singularity is recovered. As $\lambda \to \infty$,
$\beta \to 0$ and the loading is bounded: a fully transparent edge scatters
nothing and carries no suction. Between them $\beta$ falls monotonically
(figure~\ref{fig:edge}). For complex $\lambda$, as for a material with inertia,
\eqref{eq:beta} is analytic except at the branch points $\lambda = \pm\I$. These
lie on the imaginary axis and are reached only by a lossless inertive layer, at
the single frequency where $|\lambda| = 1$; any resistance, $\operatorname{Re}
\lambda > 0$, keeps $\lambda$ clear of them.

\paragraph{Compressibility does not change the exponents, and this is not
automatic.} The compressible kernel of \eqref{eq:sie} is not a Cauchy kernel.
What the argument above uses is that its non-Cauchy part $\mathcal{K}$ is
bounded (appendix~\ref{app:kernel}), so it is subdominant in the local analysis
and cannot enter $G$. The exponents are therefore fixed by the dominant singular
part alone, are independent of $k_e$, and coincide in form with the
incompressible exponents of \citet{baddoo2021}; \citet{hajian2017} reached the
corresponding steady result by the same route. The Mach number reaches the
exponents only through $\lambda$, and more weakly than
$\lambda = 2\Ad M/\beta_M$ suggests. Since $\Ad = \rho_0c_0/Z_d$ the explicit
$M$ cancels, and for the same $U$, $\omega$ and material what remains is the
Prandtl--Glauert factor, $\lambda = \lambda_0/\beta_M$, whatever the material
(\S\ref{sec:similarity}).

\paragraph{The junction exponent.}
\label{sec:junction}
At an interior point $s_j$ where $\lambda$ jumps, such as the junction between a
porous insert and the rigid remainder or between two materials, the same
argument applies with a piecewise-constant coefficient. The Riemann--Hilbert
coefficient jumps from $G_{\rm left} = \exp(2\pi\I\beta_{\rm left})$ to
$G_{\rm right} = \exp(2\pi\I\beta_{\rm right})$, and the canonical function
acquires a local exponent
\begin{equation}
\Pi \sim |\xi - s_j|^{-\delta}, \qquad
\delta = \frac{1}{2\pi\I}\log\frac{G_{\rm right}}{G_{\rm left}}
       = \beta_{\rm right} - \beta_{\rm left},
\label{eq:delta}
\end{equation}
with $\beta_{\rm left,right} = \pi^{-1}\cot^{-1}\lambda(s_j^{\mp})$. For the
common case of a porous insert ending on a rigid aft section this is
$\delta = 1/2 - \beta > 0$: moving downstream into a less permeable section
makes the loading singular, and the junction acts as the leading edge of the aft
section. Moving downstream into a more permeable section, as from a rigid
forward section onto a porous aft one, gives $\delta = \beta - 1/2 < 0$, and the
loading vanishes there as $|\xi - s_j|^{1/2-\beta}$, as at a trailing edge, as
found by \citet{baddoo2021}. For $\lambda \to 0$ on
both sides, $\delta \to 0$ and the junction disappears, which is why inserting a
junction into a rigid plate must change nothing; this property is used as a
numerical test in \S\ref{sec:verify}. For $\lambda \to \infty$ on the forward
side, $\delta \to 1/2$, the inverse-square-root singularity of a rigid leading
edge. Since $2\delta < 1$ the squared loading remains integrable and the energy
is finite.

\subsection{The permeability parameter}
\label{sec:similarity}

Because $\lambda$ is the only route by which the material enters, its parametric
dependence can be read off before \eqref{eq:sie} is solved. Two statements
follow, and both are testable.

The semi-infinite version of \eqref{eq:sie} admits the Wiener--Hopf kernel
\begin{equation}
\widetilde{K}(\chi) = \gamma(\chi) - \I\lambda\,(\chi + \kb), \qquad
\gamma(\chi) = \sqrt{\chi^2 - k_e^2},
\label{eq:kernel}
\end{equation}
in which $\chi$ is the chordwise wavenumber scaled on $b$, for the transform
$\int f(\xi)\,\E^{\I\chi\xi}\,\D\xi$, and the material enters only through
$\I\lambda$. The permeable term vanishes at $\chi = -\kb$,
the wavenumber of the frozen gust in the transformed variables, because a
disturbance convected with the stream carries no pressure jump and so drives no
seepage. This is a property of the kernel, not of the finite-chord response,
which depends on the whole kernel and on the edges. Using $k_0 = Mk_1$,
\begin{equation}
\lambda = \frac{2\I\mu}{\beta_M k_1} = \frac{\I\mu U}{\pi\beta_M f},
\qquad \mu = -\I k_0\Ad ,
\label{eq:lamu}
\end{equation}
in which the Mach number appears only through $\beta_M$. Substituting
\eqref{eq:lamu} into \eqref{eq:kernel} gives $\widetilde{K} = \gamma(\chi) +
(2\mu/\beta_Mk_1)(\chi + \kb)$, which tends to $|\chi| + 2\mu b +
(2\mu/k_1)\chi$ as $M \to 0$ at fixed $\mu$ and $k_1$. The last term is the
convective part of the boundary condition; the static porous half-plane kernel
of \citet{jaworski2013}, $|\chi| + 2\mu b$, is recovered only when it is
dropped, as at $U = 0$. The group controlling
how porous a surface appears to a gust is the ratio of the porosity parameter
$\mu$ to the \emph{hydrodynamic} wavenumber $k_1 = \omega/U$, not to the
acoustic one. For a material whose $\mu$ does not depend on frequency, such as
an inertive perforate, $|\lambda|$ falls as $1/f$, and $|\lambda| = 1$, where
the seepage and flow terms of \eqref{eq:sie} carry equal weight and
$\beta = 1/4$, defines a porous cut-off frequency $f_c = |\mu|U/(\pi\beta_M)$,
linear in speed, above which the surface appears increasingly rigid. For a
lossless perforate this is the branch point $\lambda = \pm\I$ of
\S\ref{sec:edge}; a resistive layer, whose $\mu$ is proportional to frequency,
has no cut-off.

\paragraph{What compressibility does and does not do to $\lambda$.} Write the
material law through the dimensional transfer impedance $Z_d = -\Delta p/v_s$, so
that $\Ad = \rho_0c_0/Z_d$. Then \eqref{eq:lambda} becomes
\begin{equation}
\lambda = \frac{2\rho_0U}{\beta_M Z_d} = \frac{\lambda_0}{\beta_M},
\qquad \lambda_0 = \frac{2\rho_0U}{Z_d} ,
\label{eq:lamM}
\end{equation}
and the explicit factor $M$ in \eqref{eq:lambda} is cancelled by $\Ad \propto
c_0$. When the compressible and incompressible theories are compared for the
same flow, at the same $U$, $\omega$ and material with the sound speed as the
only difference, $\lambda_0$ is therefore fixed and the whole Mach dependence of
the boundary condition is the Prandtl--Glauert factor, for any material:
$\lambda$ rises by $15\,\%$ at $M = 0.5$ and by $40\,\%$ at $M = 0.7$. That is
the comparison made in \S\ref{sec:results}. Everything else compressibility does
acts through the kernel and the Prandtl--Glauert transformation, that is through
$k_e = M\sigma/\beta_M^2$, $\kb = \sigma/\beta_M^2$ and the $1/\beta_M$ of the
quasi-steady lift. The factor $1/\beta_M$ in $\lambda$ has the same size as the
Prandtl--Glauert rise of the rigid lift and the opposite effect, since a larger
$\lambda$ means a weaker load, and \S\ref{sec:results} shows that the two cancel
in the steady, strongly permeable limit. The frequency dependence of $\lambda_0$
is set by the material. A resistive layer, $Z_d = R$, has a real $\lambda_0$
independent of frequency. An inertive perforate, $Z_d = -\I\omega m_a$ with
$m_a$ an effective mass per unit area, has $\lambda_0 = 2\I\rho_0b/(\sigma
m_a)$, imaginary and varying as $1/\sigma$. A layer with both, the form of the
material law of \citet{baddoo2021}, passes from the first behaviour to the
second as the frequency rises. If the Mach number is instead raised by raising
$U$ in the same fluid, $\lambda_0 \propto U$ for a resistive layer, and
$\lambda_0$ is fixed at fixed $\sigma$ for an inertive one.

\paragraph{Acoustic non-compactness.} The kernel carries the second compressible
parameter, $k_e = M\sigma/\beta_M^2$ of \eqref{eq:nondim}, the acoustic phase
across the half-chord. It enters through the Hankel functions of
\eqref{eq:possio}: the upwash induced at $\xi$ by the loading at $s$ arrives
with the phase of a sound wave that has crossed the distance between them, so
for $k_e|\xi - s| \gtrsim 1$ the kernels oscillate like $\E^{\I k_e|\xi - s|}$
and different parts of the chord act on one another with phase lags of order a
radian or more. The term in $k_e^2H_0^{(1)}$, which acts on the potential jump
itself, has no incompressible counterpart. Incompressible flow is the limit
$c_0 \to \infty$, in which $k_0 = 0$, the Hankel kernels reduce to the Cauchy
kernel and the retardation disappears; what remains is the hydrodynamic scale
$\kb$ of the convected gust and the wake. The two compressible parameters are
independent. The Mach number enters the boundary condition through
$\lambda_0/\beta_M$, and the forcing through the phase $\E^{\I Mk_e\xi}$ of
\eqref{eq:forcing}; $k_e$ enters the kernel and does not depend on the material
at all. At low Mach number the first is negligible and the second is not, since
$k_e \approx M\sigma$ is of order one whenever $\sigma$ is of order $1/M$
(\S\ref{sec:lowmach}).

\paragraph{Material-class scaling.} Equation~\eqref{eq:lamu} makes $|\lambda|$
depend on the material only through $\mu$, and the two classes scale oppositely.
For an inertive perforate $\mu$ is a purely geometric inverse length, $\lambda$
is imaginary, and
\begin{equation}
|\lambda| = \frac{|\mu|c}{\pi\beta_M}\bigg/\frac{fc}{U},
\label{eq:aperf}
\end{equation}
a function of the Strouhal number $fc/U$ and the perforate geometry alone. For a
resistive bulk layer $\Zt \simeq rh/(\rho_0c_0)$ carries no factor of $k_0$, so
$\mu \propto k_0$ and
\begin{equation}
|\lambda| = \frac{2\rho_0U}{\beta_M\,r\,h},
\label{eq:abulk}
\end{equation}
frequency-independent and linear in $U$; this is the parameter by which
\citet{geyer2011} organised sixteen porous aerofoils. A perforate therefore
collapses on $fc/U$ and a bulk medium on $\rho_0U/(rh)$; the two predictions are
of opposite sign and follow from the same boundary condition.

\section{Solution}
\label{sec:numerics}

\subsection{The weighted-Jacobi basis}
\label{sec:basis}

Equation~\eqref{eq:sie} is solved by collocation in a weighted-Jacobi basis
built on the exponents of \S\ref{sec:edge}. The idea of replacing the Chebyshev
(Glauert) family by Jacobi polynomials whose weights carry the
porosity-dependent exponents is due to \citet{baddoo2021}, who introduced it for
the incompressible problem and treated an interior discontinuity by splitting the
chord there into two sections, each with its own weighted-Jacobi expansion
carrying the junction exponent (their \S\,3.3). What is new here is that the
method transfers unchanged to the compressible kernel, for the reason given in
\S\ref{sec:edge}. At a junction we keep a single basis on the whole chord and
add two junction modes \eqref{eq:juncmodes} that carry the same local exponent. The basis is
\begin{equation}
g_0 = (1-\xi)^{\alpha}(1+\xi)^{-\beta}, \quad
g_n = (1-\xi)^{\alpha}(1+\xi)^{1-\beta}
      \mathcal{P}_{n-1}^{(\alpha,\,1-\beta)}(\xi), \quad
g_K = \Big(\tfrac{1+\xi}{2}\Big)^{1-\beta},
\label{eq:basis}
\end{equation}
with $\beta$ from \eqref{eq:beta} and $\alpha$ from \eqref{eq:alpha}, and the
last member representing the wake. The aerofoil modes $g_0$ and $g_n$ all carry
$(1-\xi)^{\alpha}$ and so vanish at the trailing edge; the wake mode does not
($g_K(1) = 1$), because it carries the vorticity leaving the edge. The Kutta
condition is therefore a property of the \emph{combination} the solve selects
rather than of each member separately, and it is satisfied without an extra
equation (\S\ref{sec:linsys}). At $\alpha = \beta = 1/2$ the family collapses
exactly onto Glauert's, $g_0 = \cot(\vartheta/2)$ and $g_K = \sin(\vartheta/2)$
with $\xi = -\cos\vartheta$, which is the first of the verification tests of
\S\ref{sec:verify}. At each junction two further modes are added,
\begin{equation}
q_0 = (1-\xi)^{\alpha}(1+\xi)^{1-\beta}|\xi - s_j|^{-\delta}, \qquad
q_1 = q_0\,\mathrm{sgn}(\xi - s_j),
\label{eq:juncmodes}
\end{equation}
spanning the two-sided local solution of \eqref{eq:delta}. Their envelope
vanishes at both plate edges, so the leading-edge exponent and the Kutta
condition are untouched.

\begin{figure}
\centering
\includegraphics[width=\textwidth]{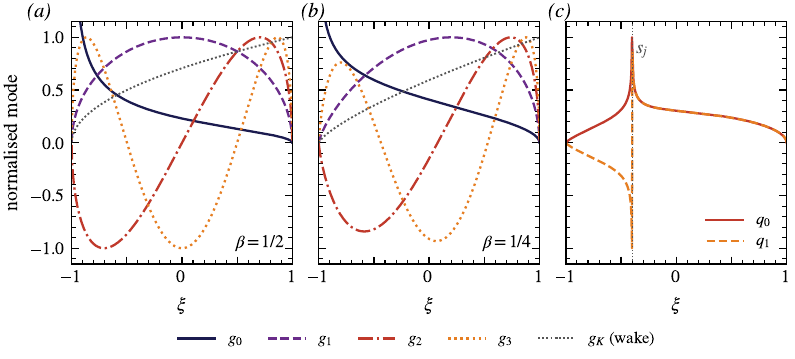}
\caption{The basis, each mode normalised by its largest value away from the
nose. ($a$) At $\beta = 1/2$ the weighted-Jacobi family reduces exactly to
Glauert's. ($b$) At $\beta = 1/4$ the nose behaviour is weaker, while the
trailing edge is unchanged.
($c$) The two junction modes \eqref{eq:juncmodes} carrying
$\delta = 1/2-\beta$. The trailing-edge exponent $\alpha$ enters the same way,
through the weight $(1-\xi)^{\alpha}$ common to every aerofoil mode.}
\label{fig:basis}
\end{figure}

Figure~\ref{fig:basis} shows what the exponents do to the basis. At
$\beta = 1/2$ the members are the classical Glauert modes, reproduced here to
$6\times10^{-13}$ (table~\ref{tab:verify}); at $\beta = 1/4$ every member is
visibly weaker at the nose while remaining identical at the trailing edge. The
difference cannot be produced by adding Glauert modes: no finite combination of
functions singular as $(1+\xi)^{-1/2}$ reproduces $(1+\xi)^{-1/4}$, which is why
the exponent must be built in rather than resolved. The same argument applies at
a porous trailing edge, where $\alpha < 1/2$, and at a junction, where the
singularity is interior.

\subsection{The linear system}
\label{sec:linsys}

\paragraph{The working unknown.} Equation~\eqref{eq:sie} is written in $\Pi$,
but $\Pi$ is a derivative combination of the jump in velocity potential across
the sheet. Let $m(\xi)$ be that jump, scaled by the same constant as $\Pi$ so
that $\Gamma = m(1)$ is the circulation. Substituting into
$\Delta p = -\rho_0(\partial_t + U\partial_{x_1})[\varphi]$ with the
$\E^{-\I\omega t}$ convention gives
\begin{equation}
\Pi(\xi) = -\frac{\partial m}{\partial\xi} + \I\kb\,m(\xi),
\label{eq:PiFromM}
\end{equation}
in which $\gamma = -\partial m/\partial\xi$ is the bound vortex-sheet strength. The two
terms carry opposite signs: the first is the quasi-steady loading a stationary
vortex sheet would carry, the second the contribution of the unsteady rate of
change of the potential jump. Their partial cancellation is why the unsteady
lift falls below its quasi-steady value as $\kb$ grows, and the Sears and
Theodorsen checks of \S\ref{sec:verify} test exactly this combination. The
potential jump is expanded in preference to $\Pi$ for three reasons.
\begin{list}{}{\settowidth{\labelwidth}{(iii)}%
  \setlength{\leftmargin}{\labelwidth}\addtolength{\leftmargin}{\labelsep}%
  \setlength{\itemsep}{2pt}}
\item[(i)] \emph{Both $m$ and $\partial m/\partial\xi$ appear.} In compressible
flow the normal velocity the sheet induces at $\xi$ has two parts: a
vortex-sheet part driven by $\partial m/\partial\xi$, which supplies the Cauchy
kernel together with an $H_1^{(1)}$ correction, and a compressibility part
driven by $m$ itself, of order $k_e^2H_0^{(1)}$. The second vanishes as
$k_e \to 0$, which is why the incompressible problem closes on the vorticity
alone and the compressible one does not.
\item[(ii)] \emph{The wake and Kelvin's theorem are statements about $\Gamma$},
which is a functional of $m$ and is carried here as an explicit unknown.
\item[(iii)] \emph{The time-domain inversion of \S\ref{sec:indicial} and the
radiation integral both act on $m$}, so solving in $m$ needs no further
differentiation of the answer.
\end{list}

\paragraph{Expansion.} The basis \eqref{eq:basis} and the junction modes
\eqref{eq:juncmodes} span $\partial m/\partial\xi$, which carries the singular
part of $\Pi$ and therefore the exponents of \S\ref{sec:edge}; the expansion of
$m$ itself
uses their antiderivatives, taken \emph{from the trailing edge},
$G_n(\xi) = -\int_\xi^1 g_n$, $G_K(\xi) = -\int_\xi^1 g_K$ and
$Q_{jk}(\xi) = -\int_\xi^1 q_k$, all of which vanish at $\xi = 1$ so that
$m(1) = \Gamma$. Then
\begin{equation}
\left.
\begin{aligned}
m(\xi) &= \Gamma + \sum_{n=0}^{N} c_n G_n(\xi) + c_K G_K(\xi)
        + \sum_j\big(c_{j0}Q_{j0}(\xi) + c_{j1}Q_{j1}(\xi)\big),\\[2pt]
\frac{\partial m}{\partial\xi} &= \sum_{n=0}^{N} c_n g_n(\xi) + c_K g_K(\xi)
        + \sum_j\big(c_{j0}q_0(\xi;s_j) + c_{j1}q_1(\xi;s_j)\big),
\end{aligned}\;\right\}
\qquad c_K = \I\kb\,\Gamma ,
\label{eq:expand}
\end{equation}
and $\Pi$ follows pointwise from \eqref{eq:PiFromM}. The wake coefficient is not
independent: the wake carries $m_w(\xi) = \Gamma\E^{\I\kb(\xi-1)}$ for
$\xi > 1$, so continuity of $\partial m/\partial\xi$ across $\xi = 1$, which means no
concentrated vortex at the edge, fixes $c_Kg_K(1) = c_K = \I\kb\Gamma$. That
single tie makes the Kutta condition automatic rather than imposed, because
\eqref{eq:PiFromM} then gives
\begin{equation}
\Pi(1) = -c_K + \I\kb\Gamma = 0
\label{eq:kutta}
\end{equation}
identically, for every $\Gamma$, every porosity and every trailing-edge exponent
$\alpha > 0$. The $G_n$ are available in closed form through a Jacobi identity;
the $Q_{jk}$ are not, because $q_k$ carries the interior algebraic singularity,
and they are evaluated once per junction by cumulative quadrature on a graded
grid. The unknown vector is $\{c_0,\dots,c_N,\Gamma,c_{j0},c_{j1}\}$:
$N + 2 + 2n_j$ numbers for $n_j$ junctions, with $N \simeq 1.6\kb + 2k_e + 12$
set automatically and raised where the admittance distribution has interior
kinks (\S\ref{sec:verify}).

\paragraph{Collocation and closure.} Substituting \eqref{eq:expand} into
\eqref{eq:sie} and enforcing it at $n_c = N+1+2n_j$ points
$\xi_i = -\cos\vartheta_i$, $\vartheta_i = (i-\tfrac12)\pi/n_c$, gives $n_c$
rows; the variable coefficient enters simply as $\lambda(\xi_i)$ on the
diagonal, which is why an arbitrary $\lambda(\xi)$ costs nothing. Kelvin's
circulation theorem supplies the one remaining row. For a closed material
contour that at $t \to -\infty$ lay in undisturbed fluid and encloses the
aerofoil together with all the vorticity it has shed, the circulation cannot
change \citep{batchelor1953}, so
\begin{equation}
\underbrace{\int_{-1}^{1}\frac{\partial m}{\partial\xi}\,\D\xi}_{\text{bound}}
+ \underbrace{(-\Gamma)}_{\text{shed}} = 0
\qquad\Longleftrightarrow\qquad m(-1) = 0 .
\label{eq:kelvin}
\end{equation}
The two auxiliary conditions act at opposite ends of the plate and play distinct
roles: Kutta fixes the \emph{behaviour} of the solution at $\xi = +1$, through
the basis; Kelvin fixes the \emph{amount} of circulation at $\xi = -1$, through
the one extra row. The result is a square $(N+2+2n_j)$ system, solved directly.

Two consequences follow. First, the cost is that of the classical rigid problem:
the same number of unknowns, the same quadrature, one dense solve. Porosity
changes the exponents in the basis functions and the diagonal, not the size or
the structure of the matrix. Second, a basis without the right exponents fails
in a specific way: omitting \eqref{eq:juncmodes} removes the only columns able
to represent $|\xi - s_j|^{-\delta}$, and the remaining columns then fit the
collocation data as well as they can, so the error stops falling well short of
zero, at a level set by the missing function. That is the $1$--$5\,\%$ stall
found in \S\ref{sec:verify}, and it is a statement about the span of the basis
rather than about the number of modes.

\subsection{Quadrature}
\label{sec:quad}

No closed-form Hilbert transform of the weighted-Jacobi family is needed. The
Cauchy singularity is removed by subtraction,
\begin{equation}
\pvint_{-1}^{1}\frac{g(\xi)}{s-\xi}\,\D\xi
= \int_{-1}^{1}\frac{g(\xi)-g(s)}{s-\xi}\,\D\xi + g(s)\log\frac{1+s}{1-s},
\label{eq:subtract}
\end{equation}
after which Gauss--Legendre panels graded towards the singular points suffice.
The panels must additionally be subdivided so that no panel spans more than
$\Delta\vartheta = \pi n_p/(4N)$, with $n_p$ Gauss points per panel, otherwise
the $\cos(N\vartheta)$ oscillation of the high-order modes is aliased; without
this refinement the junction-invariance test of \S\ref{sec:verify} returns a
$7\,\%$ error. Details are in appendix~\ref{app:quad}.

\subsection{Indicial responses}
\label{sec:indicial}

The Wagner and K\"ussner functions are the time-domain counterparts of the
Theodorsen and Sears functions, and they are what an aeroelastic, gust-load or blade--vortex-interaction calculation
actually requires. They are obtained here
by inverting the computed transfer function rather than by solving a separate
problem. Let $L(\sigma)$ be the lift response to the harmonic forcing of
\eqref{eq:forcing} at reduced frequency $\sigma$, normalised on its quasi-steady
value. The indicial response to a step of the same kind is
\begin{equation}
\phi(\tau) = \frac{2}{\pi}\int_{0}^{\infty}
   \frac{\operatorname{Re}L(\sigma)}{\sigma}\,\sin(\sigma\tau)\,\D\sigma ,
\label{eq:indicial}
\end{equation}
$\tau = Ut/b$ being semichords travelled, the standard inversion
\citep{bisplinghoff1955} valid because $L$ is the transfer function of a causal,
stable, real system. Two properties of the present formulation matter for it.
The gust phase is referred to the leading edge \eqref{eq:forcing}, so the
K\"ussner response starts at $\tau = 0$ rather than at $\tau = -1$; and the
apparent-mass part of the heave response, which is an impulse at $\tau = 0$ and
$-\I\sigma/2$ in the rigid limit, is removed before inverting, since
Wagner's function describes what is left. Both the subtraction and the tail of
\eqref{eq:indicial} beyond the computed range are given in
appendix~\ref{app:indicial}; the end-to-end accuracy of the scheme is the
$0.007$ with which it reproduces the classical Wagner function
(\S\ref{sec:indicialres}).

\section{Verification}
\label{sec:verify}

Table~\ref{tab:verify} lists the verification tests. They fall into three
groups: tests of the machinery that the rigid theory shares (the first six
rows), tests of the porous terms against closed forms and against the exponents
they are built on (the next eight), and tests of the assembled solver against
independent implementations (the next five); the last row is a consistency
check made on every solve. The sixth row is the only classical check made at
finite Mach number: at high reduced frequency the loading of a heaving plate
approaches the acoustic value $\Delta p = 2\rho_0c_0w$, which is
$2\beta_M/(\pi M)$ times the compressible quasi-steady lift. The test is
sensitive to the Prandtl--Glauert factor of \eqref{eq:forcing} and
\eqref{eq:lift}: omitted from both, it returns $1.28$ at $M = 0.5$ in place of
$1.10$. For a permeable plate the same argument, with flow through the material
as a second path in parallel with the piston motion, gives the limit
\eqref{eq:hflimit} of \S\ref{sec:searsmach}, which the solver reproduces to $10^{-4}$ at $\sigma = 50$.

\begin{table}
\centering\small
\begin{tabular}{@{}lll@{}}
\toprule
test & reference & error \\
\midrule
Basis at $\alpha=\beta=1/2$ & $\cot(\vartheta/2)$, $\sin(\vartheta/2)$ & $5.8\times10^{-13}$ \\
Possio symbol, $|k_e| = 0.5$--$10$ & $\tfrac12\I\sqrt{k_e^2-\nu^2}$, closed form & $7.3\times10^{-16}$ \\
\quad the same, as assembled in \S\ref{sec:numerics} & the same, by quadrature & $1.7\times10^{-5}$ \\
Sears function, $\sigma = 0.1$--$8$ & $[J_0+\I J_1]\mathcal{C}(\sigma) - \I J_1$ & $4.3\times10^{-5}$ \\
Heave response, $\sigma = 0.02$--$8$ & $\mathcal{C}(\sigma) - \I\sigma/2$ (incl.\ apparent mass) & $1.4\times10^{-4}$ \\
Heave, $\sigma = 40$--$50$, $M = 0.3$--$0.7$ & piston limit, and \eqref{eq:hflimit} for $\lambda_0 = 1$--$30$ & $2\times10^{-3}$ rel. \\
Steady lift, $\lambda = 0.02$--$200$ & $C_L = 4\pi\alpha_i\beta$, closed form \eqref{eq:CL} & $2.7\times10^{-4}$ \\
Seepage drag & $C_{d,\rm seep} = \alpha_i C_L$, closed form \eqref{eq:Cd} & $4.0\times10^{-4}$ \\
Porous quasi-steady, $\lambda = 0$--$8$ & $2\beta$, from the \emph{unsteady} solver & $8.9\times10^{-4}$ \\
Junction invariance on a rigid plate & must be exactly unity & $3.5\times10^{-10}$ \\
Edge exponent read off the solution & $\beta$ \eqref{eq:beta}, $\lambda = 0$--$8$, $k_e \le 6.7$ & $3\times10^{-3}$ rel. \\
Junction exponent read off the solution & $\delta$ \eqref{eq:delta}, $M = 0$ and $0.5$ & $8\times10^{-3}$ rel. \\
Generalised basis at $\alpha = 1/2$ & the rigid-trailing-edge basis & bit-identical \\
Mode count, $\sigma = 10$, $M = 0.7$ & $N$ default against $N = 160$ & $<10^{-5}$ \\
Steady lift, $\lambda = 0$--$10$ & \citet{baddoo2021} eq.~(3.1), solved independently & $6.5\times10^{-5}$ \\
Unsteady, $k_e \to 0$, $k = 0.01$--$10$ & \citet{baddoo2021}, their published code & $0.09\,\%$ \\
Rigid, $M = 0.3$--$0.7$, $\sigma = 0.2$--$10$ & physical-variable solution, appendix~\ref{app:indep} & $8\times10^{-4}$ \\
Permeable, $\lambda_0 = 1, 3$, $\sigma = 0.5$--$10$ & the same & $8\times10^{-6}$ \\
Independent reimplementation & \textsc{matlab} vs \textsc{python} & $3.1\times10^{-10}$ \\
Loading/potential identity & \eqref{eq:PiFromM}, evaluated on every solve & $5.0\times10^{-4}$ \\
\bottomrule
\end{tabular}
\caption{Verification tests. $\alpha_i$ is the incidence, not the trailing-edge
exponent. $\mathcal{C}(\sigma) = H_1^{(1)}(\sigma)/[H_1^{(1)}(\sigma) - \I
H_0^{(1)}(\sigma)]$ is Theodorsen's function in the $\E^{-\I\omega t}$
convention, $J_n = J_n(\sigma)$, and the Sears phase is referred to midchord.
The second row is the most sensitive test of the quadrature; the last
is the only one available at no cost on every solve.}
\label{tab:verify}
\end{table}

\begin{figure}
\centering
\includegraphics[width=0.72\textwidth]{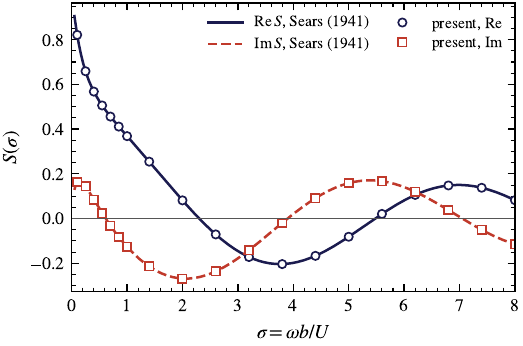}
\caption{The rigid, incompressible limit against the Sears function. Lines,
the classical function \citep{sears1941}; symbols, the present solver at
$k_e \to 0$, with the phase referred to midchord. The difference is below
$5\times10^{-5}$ over $\sigma = 0.1$--$8$.}
\label{fig:sears}
\end{figure}

\paragraph{The compressible operator.} All the classical checks except the
piston limit are made at $k_e \to 0$, where the $k_e^2H_0^{(1)}$ term and the
regularised $H_1^{(1)}$ remainder vanish, so they do not test the compressible
part of the operator. That part is verified directly, through its Fourier
symbol: on an unbounded sheet, a potential jump $m = \E^{\I\nu s}$ must induce
the normal velocity $w = \tfrac12\I r\,m$, with $r = (k_e^2-\nu^2)^{1/2}$ and
$\operatorname{Im} r \ge 0$, which is the exact response of the Helmholtz
problem \eqref{eq:helm}. Applying \eqref{eq:possio} to $m = \E^{\I\nu s}$, with
$\gamma = -\I\nu m$ and the transforms
\begin{gather*}
\int_{-\infty}^{\infty}\!\! H_0^{(1)}(k_e|u|)\,\E^{-\I\nu u}\D u = \frac{2}{r},
\qquad
\int_{-\infty}^{\infty}\!\! H_1^{(1)}(k_e|u|)\,\mathrm{sgn}(u)\,\E^{-\I\nu u}\D u
 = \frac{-2\I\nu}{k_e r},
\end{gather*}
gives $(-\I\nu^2 + \I k_e^2)/(2r) = \tfrac12\I r$, as required. Because the
identity must hold for every $\nu$, it tests each Hankel term and their relative
coefficient separately, which a comparison with a tabulated lift coefficient
would do only indirectly. It is satisfied to $7\times10^{-16}$ in closed form,
and to $1.7\times10^{-5}$ by quadrature over $|k_e| = 0.5$--$10$ and
$\nu = 0.1$--$8$, both directly and through the split into a Cauchy principal
value, the regularised $H_1^{(1)}$ and $k_e^2H_0^{(1)}$ that
\S\ref{sec:numerics} assembles. End to end, the solver approaches the classical
Sears function continuously as $k_e \to 0$ at fixed $\kb$: the difference falls
from $5.5\times10^{-2}$ at $k_e = 0.1$ to $1.3\times10^{-3}$ at $k_e = 10^{-2}$,
at the $k_e^2\log k_e$ rate of the small-argument expansion of the kernel, and
reaches the solver's own floor of $4\times10^{-5}$ by $k_e = 10^{-3}$.
Figure~\ref{fig:sears} shows the resulting agreement with the Sears function
over $\sigma = 0.1$--$8$.

\paragraph{An independent solution at finite Mach number.} The Fourier-symbol
test verifies the operator and the $k_e \to 0$ tests verify its limit, but
neither tests the assembled compressible solution, and the Prandtl--Glauert
factor of \eqref{eq:forcing} and \eqref{eq:lift}, which is invisible at $M = 0$,
has so far been tested only in the high-frequency heave limit. The assembled
solution is therefore compared with a second solution of the same physical
problem (appendix~\ref{app:indep}). Its unknown is the physical pressure jump
rather than the transformed potential, it has no wake and no Prandtl--Glauert
transformation, and the kernel is applied through its exact Fourier symbol in
physical variables; only the edge exponents of \S\ref{sec:edge} are common to
the two methods. The absolute difference in the lift is at most
$8\times10^{-4}$ for a rigid plate over $M = 0.3$--$0.7$ and
$\sigma = 0.2$--$10$, and at most $8\times10^{-6}$ for permeable plates with
$\lambda_0 = 1$ and $3$ over $\sigma = 0.5$--$10$, in magnitude and phase and
for both the gust and the heave forcing. The larger rigid-plate difference comes
from truncating the wavenumber integral of the second method
(appendix~\ref{app:indep}), not from the first.

\paragraph{The incompressible limit: comparison with \citet{baddoo2021}.} Setting $k_e \to 0$ in
\eqref{eq:sie} removes the $k_e^2H_0^{(1)}$ terms and the Possio kernel
collapses to the Cauchy kernel alone, so the equation becomes the incompressible
singular integral equation of \citet{baddoo2021} with $\lambda$ in the role of
their porosity parameter; the exponents \eqref{eq:beta}, \eqref{eq:alpha} and
\eqref{eq:delta} are independent of $k_e$ and so are already common to both.
Their results are therefore the $k_e \to 0$ limit of the present solver, and the
results of \S\ref{sec:results} are its continuation in $k_e$. That claim is
checked in three increasingly demanding ways.

First, structurally, in the steady limit: their equation~(3.1) was discretised
from scratch in their variable, the bound vorticity, and in their basis,
sharing no code with \S\ref{sec:numerics}, and its circulation reproduces
$\Gamma(\psi)/\Gamma(0) = 2\beta$ to $6.5\times10^{-5}$.

Second, in the unsteady quasi-steady limit, where the solver is asked for limits
it was not constructed to satisfy: with $\lambda = 0$ a heave forcing returns
$\mathcal{C}(\sigma) - \I\sigma/2$ to $3.2\times10^{-4}$ over $\sigma \le 5$, so
the circulatory \emph{and} non-circulatory parts of the classical response are
both recovered; with $\lambda > 0$ the quasi-steady limit of the unsteady
solution returns $2\beta$ to $8.9\times10^{-4}$ over $\lambda = 0$--$8$, for
heave and gust forcing alike.

\begin{figure}
\centering
\includegraphics[width=\textwidth]{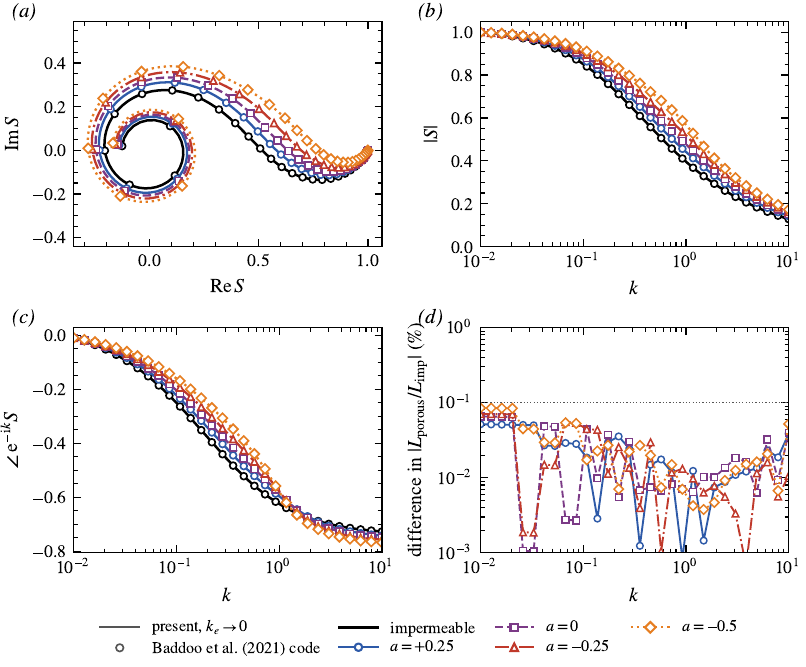}
\caption{The present solver at $k_e \to 0$ (lines) against the published code
of \citet{baddoo2021} (open symbols), on the porosity profiles of their
figure~10: an impermeable leading edge, $\Psi = 1/\Phi$ rising linearly from
zero at $x = a$ to $0.5$ at $x = a + 0.8$, $\rho_e = 1.5$; black, impermeable.
Each curve is normalised at $k = 10^{-2}$, as their figure is. ($a$) Argand
diagram, ($b$) magnitude and ($c$) phase of the porous Sears function, laid out
as their figure~10. ($d$) The relative difference between the two codes in the
ratio $|L_{\rm porous}/L_{\rm impermeable}|$, which is independent of
normalisation and time convention; the dotted line marks $0.1\,\%$. Both codes
at their converged basis size: $N = 60$ here, $N = \lfloor 25 +
20\sqrt{k}\rfloor$ in theirs.}
\label{fig:bdcode}
\end{figure}

Third, and most stringently, against their published code itself
(\url{https://github.com/baddoo/unsteady-porous-aerofoils}), run on the porosity
profiles exactly as their figure~10 is generated: an impermeable leading edge,
$\Psi = 1/\Phi$ ramping linearly to $0.5$ over $0.8$ of the chord from
$a \in \{0.25, 0, -0.25, -0.5\}$ and constant thereafter, so that the trailing
edge is porous, and $\rho_e = 1.5$. Their $\psi$ is written for the time
convention $\E^{+\I kt}$, so in the present convention the same material is its
complex conjugate; the porosity supplied to the present solver and the lift it
returns were both converted accordingly before the two codes were compared.
Figure~\ref{fig:bdcode} shows the result. The ratio $|L_{\rm porous}/L_{\rm impermeable}|$, which is free of any
normalisation or phase convention, agrees with their code to within
$0.09\,\%$ on all four profiles over $k = 0.01$--$10$, with no growth in $k$
($0.05$, $0.07$, $0.07$ and $0.09\,\%$ for the four ramps); the impermeable
control agrees to $2.4\times10^{-5}$. This is
point-by-point agreement with an independent implementation of the limiting
case, on configurations chosen by its authors rather than by us, and it
includes a \emph{porous trailing edge}, $\alpha \approx 0.15$, which exercises
\eqref{eq:alpha}. The features \citet{baddoo2021} report are reproduced with
it: $|S|$ grows with permeable length, and the porous phase curves cross the
impermeable one, here at $k \approx 1.4$--$2.2$, the crossover moving to lower
frequency as the permeable length grows (figure~\ref{fig:bdcode}$c$). On the
impermeable curve the present solver reproduces their closed form (6.3) to
$4.6\times10^{-6}$.

\begin{figure}
\centering
\includegraphics[width=\textwidth]{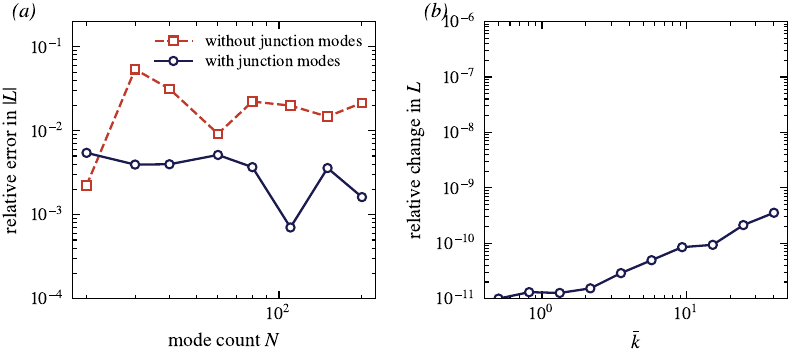}
\caption{($a$) Convergence of the lift with and without the junction modes
\eqref{eq:juncmodes}, for a permeable insert over the forward $30\,\%$ of the
chord, $\lambda = 3$, $\sigma = 2$; the reference is the same solver at
$N = 280$. ($b$) The complementary test: a junction placed on a rigid plate,
where $\delta = 0$, must change nothing.}
\label{fig:conv}
\end{figure}

\paragraph{Convergence and the junction modes.} The decisive convergence result
concerns the junction modes (figure~\ref{fig:conv}). For a permeable insert with
$\lambda = 3$ over the forward $30\,\%$ of the chord at $\sigma = 2$, the lift
computed \emph{without} the junction modes wanders at the $1$--$5\times10^{-2}$
level and does not settle as $N$ is raised from $20$ to $200$: the expansion
converges, but not to the right answer, because a global basis cannot represent
an algebraic singularity at an interior point. \emph{With} them the error
settles at a few times $10^{-3}$, an order of magnitude better and still
falling, and the remaining level is the price of representing $|\xi-s_j|^{-\delta}$
with two modes rather than a local expansion. Raising the quadrature resolution
of the junction modes by a factor of six changes nothing, so this is a basis
limit and not a quadrature one. The asymmetry between the two curves is the
numerical statement of \eqref{eq:delta}: a basis that does not carry the
junction exponent cannot converge to the solution, which is why
\citet{baddoo2021} split the chord at a discontinuity and why the junction
modes are added here. The complementary test is exact rather than approximate,
and it is the sharpest test of the quadrature: a junction placed on a
\emph{rigid} plate, where $\delta = 0$ and the modes must vanish, must change
nothing, so any residual is discretisation error. It changes the lift by
$3.5\times10^{-10}$ over $\kb = 0.5$--$40$. A \emph{smoothly varying}
$\lambda(\xi)$ with interior kinks, such as the ramps of
figure~\ref{fig:bdcode}, converges more slowly than a piecewise-constant insert
with explicit junction modes, so the default mode count must be raised there.
$N = 60$ suffices over $k \le 10$, and raising the count from the default to
$60$ takes the comparison of figure~\ref{fig:bdcode} from $0.63\,\%$ to
$0.09\,\%$. The remaining difference is at the level expected of two
independent spectral discretisations rather than a property of either.

\section{Results}
\label{sec:results}

All results below are for a uniformly permeable plate or for a permeable insert
of length $\ell$ at the leading edge, with $\lambda$ quoted directly rather than
through a particular material, so that they apply to any surface with the same
$\lambda$. Unless stated otherwise the surface is a resistive layer: $\lambda_0
= 2\rho_0U/R$ is real and independent of frequency, and the compressible and
incompressible responses are compared at the same $U$, $\omega$ and $R$, so that
$\lambda = \lambda_0/\beta_M$ by \eqref{eq:lamM} and the sound speed is the only
difference. A layer with inertia as well as resistance is considered in
\S\ref{sec:inertive}. Reduced frequency is the classical
$\sigma = \omega b/U$ throughout. Loads at finite Mach number are reported on a
single basis, the incompressible quasi-steady lift of the rigid plate,
$2\pi\rho_0Uw_0b$ per unit span, rather than on the compressible one,
$2\pi\rho_0Uw_0b/\beta_M$. The Prandtl--Glauert rise is therefore kept, so that
a rigid plate at low frequency reads $1/\beta_M$ rather than unity, and a change
with Mach number is a change in the load itself.

\subsection{The loading at a permeable edge}
\label{sec:loading}

\begin{figure}
\centering
\includegraphics[width=\textwidth]{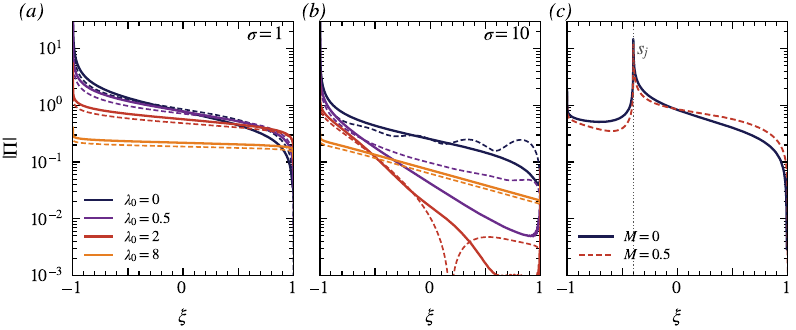}
\caption{Chordwise loading, uniformly permeable plate, at ($a$) $\sigma = 1$ and
($b$) $\sigma = 10$; solid $M = 0$, dashed $M = 0.5$ with the same material, so
that $\lambda = \lambda_0/\beta_M$; colour and the legend give $\lambda_0$.
($c$) A permeable insert over the forward
$30\,\%$ of the chord, $\lambda = 3$, $\sigma = 2$: the nose singularity is
weakened and a junction singularity $|\xi-s_j|^{-\delta}$ appears at the dotted
line.}
\label{fig:loading}
\end{figure}

Figure~\ref{fig:loading} shows the chordwise loading. Three things in it are
worth separating, because they are usually run together.

\emph{The exponents are in the solution, not only in the basis.} Fitting
$|\Pi| \sim (1+\xi)^{-\beta}$ over the three decades $10^{-5} \le 1+\xi \le 10^{-2}$
returns $\beta = 0.5002$, $0.3527$, $0.1478$ and $0.0396$ for
$\lambda = 0$, $0.5$, $2$ and $8$, against $0.5000$, $0.3524$, $0.1476$ and
$0.0396$ from \eqref{eq:beta}, which is agreement to three figures; at $\sigma =
10$ and $M = 0.5$, where $k_e = 6.7$, the agreement is still within $2\,\%$. That the fit reproduces
\eqref{eq:beta} at finite $k_e$ is the numerical counterpart of the analytical
statement of \S\ref{sec:edge}: the compressible part of the kernel is
subdominant at the edge and does not touch the exponent. At the junction of a
finite insert the same fit returns $\delta = 0.3943$ against $0.3976$ from
\eqref{eq:delta} at $M = 0$, and $0.4096$ against $0.4105$ at $M = 0.5$.

\emph{The same material gives a weaker singularity at higher Mach number.}
Because $\lambda(M) = \lambda_0/\beta_M$, a surface with $\lambda_0 = 2$
presents $\lambda = 2.31$ at $M = 0.5$, and the fitted exponent falls from
$0.148$ to $0.130$. Nothing about the material has changed; the exponent has,
by $12\,\%$. A rigid edge has a Mach-independent singularity and a permeable one
does not, which is a statement no static permeability condition can make.

\emph{The loading redistributes as much as it falls.} Panels ($a$) and ($b$)
are plotted logarithmically because the interesting comparison is not the peak
but the shape: as $\lambda$ rises, the nose loading collapses while the aft
loading is left almost untouched, so the centre of pressure moves aft and the
lift falls by less than the edge loading does. The finite insert in panel ($c$)
makes the mechanism explicit. The insert weakens the nose, but the junction
then carries $|\xi-s_j|^{-\delta}$ and behaves as a second, weaker leading edge. That is why a treated nose does not
simply remove the edge but moves a reduced version of it downstream. At $\ell/c = 0.3$, $\lambda = 3$ and
$\sigma = 2$ the lift is $0.208$ of the rigid quasi-steady value against
$0.279$ for a rigid plate at the same $\sigma$, a $25\,\%$ reduction for a
treatment covering $30\,\%$ of the chord.

\subsection{The porous Sears function at finite Mach number}
\label{sec:searsmach}

\begin{figure}
\centering
\includegraphics[width=\textwidth]{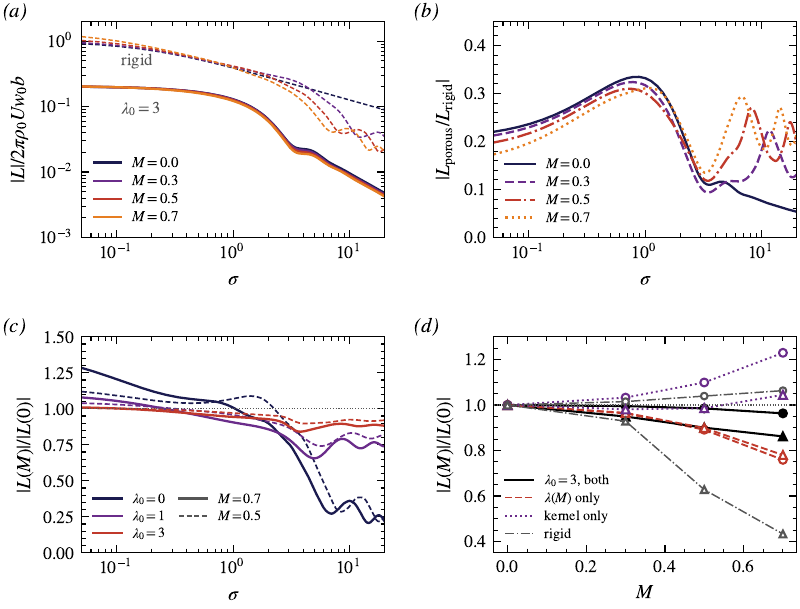}
\caption{The porous Sears function at finite Mach number for a resistive layer,
compared at the same flow speed, frequency and material so that $\lambda =
\lambda_0/\beta_M$. Every load is referred to the
incompressible quasi-steady lift of the rigid plate. ($a$) $|L|$ for
$\lambda_0 = 3$ (solid) and for a rigid plate (dashed), coloured by Mach
number. ($b$) The fraction of the rigid load that survives the treatment; the
$M = 0$ curve is what an incompressible theory would predict. ($c$) The Mach
factor $|L(M)|/|L(0)|$ for $\lambda_0 = 0$, $1$ and $3$, at $M = 0.7$ (solid) and
$0.5$ (dashed). ($d$) The two channels at $\lambda_0 = 3$ against Mach number:
the full response (filled symbols), the response with only $\lambda(M)$
varying, the response with only the kernel varying, and the rigid plate, for
which only the kernel acts; circles $\sigma = 0.5$, triangles $\sigma = 5$.}
\label{fig:searsmach}
\end{figure}

What is held fixed as the Mach number varies is the flow speed, the frequency
and the material, so that $\lambda = \lambda_0/\beta_M$ by \eqref{eq:lamM} while
$k_e = M\sigma/\beta_M^2$ grows from zero. Figure~\ref{fig:searsmach} is the
result, and it separates cleanly into what is classical and what is not.

\paragraph{The rigid response is the classical one.} Compressibility raises the
rigid load at low frequency by the Prandtl--Glauert factor, $1/\beta_M = 1.15$
at $M = 0.5$ and $1.40$ at $M = 0.7$, and collapses it once the chord is no
longer acoustically compact: at $\sigma = 10$ the load at $M = 0.5$ is $0.32$ of
its incompressible value, and at $\sigma = 20$ it is $0.25$ (panel $a$, dashed).
Above $\sigma \approx 5$ the collapse carries shallow interference minima, at
$\sigma \approx 12.4$ for $M = 0.3$, $8.8$ and $18$ for $M = 0.5$, and $7.1$ and
$14.8$ for $M = 0.7$, spaced so that $2k_0b/(1+M)$, the phase a sound wave
accumulates in crossing the chord with the stream, increases by very nearly
$2\pi$ from one to the next. None of this is new. It is the behaviour of
Possio's problem that the similarity rules of \citet{graham1970}, the
approximate theories of \citet{osborne1973}, \citet{amiet1976} and
\citet{kemp1976}, and the compressible indicial functions used in aeroelastic and rotor
analysis \citep{leishman1988} all describe; the present solver reproduces it,
and the independent solution of appendix~\ref{app:indep} reproduces the minima
to $2\times10^{-3}$. It is the reference against which the permeable results are
read.

\paragraph{The permeable response depends on Mach number through its
permeability.} Panel ($c$) shows the Mach factor $|L(M)|/|L(0)|$ for
$\lambda_0 = 0$, $1$ and $3$, and further cases at $\lambda_0 = 0.5$ and $8$
complete the picture. At $M = 0.7$ and $\sigma = 5$ the load is $0.43$ of its incompressible
value for a rigid plate, $0.53$ at $\lambda_0 = 0.5$, $0.66$ at $\lambda_0 = 1$,
$0.86$ at $\lambda_0 = 3$ and $0.95$ at $\lambda_0 = 8$: compressibility changes
the load of a weakly to moderately permeable surface by between a tenth and a
half, and that of a very permeable one hardly at all. At $\lambda_0 = 3$ the
three compressible curves of panel ($a$) lie close to the incompressible one;
over $M \le 0.7$ and $\sigma \le 20$ the load changes by at most $16\,\%$ (near
$\sigma = 3.8$ at $M = 0.7$), and by $1$--$2\,\%$ in the quasi-steady limit,
where the rigid load changes by $40\,\%$. The sensitivity falls steadily as the
surface becomes more permeable at high frequency and in the quasi-steady limit,
where the same ratio is available in closed form (\S\ref{sec:steady}) and
interpolates between the Prandtl--Glauert factor of the rigid plate and unity.
Compressibility therefore matters for a permeable surface, but by an amount and
with a frequency dependence that neither the rigid compressible theory nor the
incompressible porous theory supplies.

\paragraph{Why: the load of a permeable surface is set by the material.} In the
strongly permeable limit the surface simply passes the imposed upwash: the
disturbance potential vanishes, the seepage velocity in \eqref{eq:bcconv}
balances the upwash, $|v_s| = |w|$, and the pressure jump is whatever the
material needs to drive it through. By \eqref{eq:vs},
\begin{equation}
|\Delta p| \to \frac{\rho_0c_0}{|\Ad|}\,|w| = \frac{2\rho_0U}{|\lambda_0|}\,|w| ,
\label{eq:seep}
\end{equation}
the second form following from \eqref{eq:lambda} with $\lambda = \lambda_0/\beta_M$.
The flow around the surface does not enter, and with it neither does
compressibility: \eqref{eq:seep} contains no $M$. On the common basis of
\S\ref{sec:results}, the incompressible quasi-steady lift of the rigid plate, it
gives $|L| \to (2/\pi\lambda_0)\,|\sin\sigma/\sigma|$ for the gust
(figure~\ref{fig:searsmach}) and $2/(\pi\lambda_0)$ for heave
(\S\ref{sec:theo}), at every Mach number. How the
limit is approached is where compressibility enters, and for heave at high
frequency it can be written down. Each face of the plate then radiates as a
piston, with pressure $\rho_0c_0$ times its normal velocity. The imposed upwash
is taken up partly by this motion and partly by flow through the material, both
driven by the same pressure jump, so the acoustic and material admittances add:
eliminating the face velocity between the piston relation and
\eqref{eq:bcconv} gives
\begin{equation}
|\Delta p| \to \frac{2\rho_0U}{\lambda_0 + M}\,|w| ,
\qquad
\frac{|L|}{2\pi\rho_0Uw_0b} \to \frac{2}{\pi(\lambda_0 + M)} ,
\label{eq:hflimit}
\end{equation}
which is the piston value $2/(\pi M)$ for a rigid plate and the seepage limit as
$M/\lambda_0 \to 0$. The argument is local and holds for any material law, with
$\lambda_0$ evaluated at the frequency in question: for a resistive layer
$\lambda_0$ is constant and the load tends to a plateau, whereas for an inertive
one $\lambda_0 \to 0$ and the load tends to the piston value of a rigid plate
(\S\ref{sec:inertive}). The solver reproduces \eqref{eq:hflimit} to four figures at
$M = 0.5$ and $\sigma = 50$, and also at $M = 0.7$ and $\sigma = 20$ for
$\lambda_0 = 3$, $8$ and $30$, where the load lies $M/(\lambda_0 + M) = 19$, $8$
and $2\,\%$ below the seepage limit. The high-frequency Mach sensitivity is
therefore of first order in $M/\lambda_0$, whereas the steady one
(\S\ref{sec:steady}) is of second order. For the gust the leading term is
approached more slowly as $\sigma$ grows, since the edges then carry more of
the load, but its Mach independence is inherited: at $\lambda_0 = 30$ the gust
load changes by at most $1.4\,\%$ between $M = 0$ and $0.7$. The same limit explains
the shoulder in every permeable curve near $\sigma = \pi$, where the chord
average of the gust, and with it the leading term, vanishes. A rigid plate is
the opposite extreme: its load is set entirely by the flow around it, which is
where compressibility acts.

\paragraph{The two channels act in opposite directions.} Panel ($d$) separates
them at fixed $\sigma$, on the same basis. The boundary-condition channel, in which $\lambda$ rises as $1/\beta_M$ while
the kernel is held incompressible, lowers the permeable load by $22$--$24\,\%$
at $M = 0.7$, almost independently of $\sigma$. The kernel channel, which
carries the Prandtl--Glauert rise, raises it by $23\,\%$ at $\sigma = 0.5$ and
by $5\,\%$ at $\sigma = 5$; at $\sigma = 5$ the same kernel, acting on a rigid
plate, lowers the load by $57\,\%$. At $\lambda_0 = 3$ the two cancel in part.
The full change is $4\,\%$ at $\sigma = 0.5$, $7\,\%$ at $\sigma = 2$ and
$14\,\%$ at $\sigma = 5$, where the kernel's rise has faded, and the product of
the two channels reproduces the full response to within $4\,\%$ for $\sigma \le
2$ and $6\,\%$ at $\sigma = 5$. The cancellation improves as the
surface becomes more permeable: the full change is $2$--$5\,\%$ at
$\lambda_0 = 8$ over $\sigma \le 20$, and it is exact in the steady, strongly
permeable limit (\S\ref{sec:steady}).

\paragraph{What the treatment is worth.} The surviving fraction
$|L_{\rm porous}/L_{\rm rigid}|$, the load of the treated plate divided by that
of the rigid plate at the same $\sigma$ and $M$, is shown in panel ($b$). It is a
ratio whose numerator hardly depends on Mach number and whose denominator
carries the whole classical collapse. It
therefore inherits that collapse. At low reduced frequency the incompressible
theory is pessimistic: it gives $0.236$ at $\sigma = 0.1$ where $M = 0.5$ gives
$0.217$, and $0.332$ against $0.301$ at $\sigma = 1$. The largest difference is
the quasi-steady one: the compressible fraction is then the incompressible one
times $2\beta(\lambda_0/\beta_M)/2\beta(\lambda_0)$, which is $0.87$ at $M =
0.5$ and $0.73$ at $M = 0.7$. The curves cross near
$\sigma = 2.6$ at $M = 0.5$ ($1.5$ at $M = 0.7$, $4.9$ at $M = 0.3$), and above
that the incompressible theory is increasingly optimistic: at $\sigma = 10$ it predicts that $7.6\,\%$ of the rigid
load survives where $M = 0.5$ gives $21.6\,\%$. Over the non-compact range
$\sigma = 5$--$20$, it underestimates the surviving fraction by a factor of
between $1.4$ and $4.2$ at $M = 0.5$ and between $2.0$ and $4.3$ at $M = 0.7$.
The peaks of the compressible curves above $\sigma = 5$ sit just below the
interference minima of the rigid response.

\begin{figure}
\centering
\includegraphics[width=\textwidth]{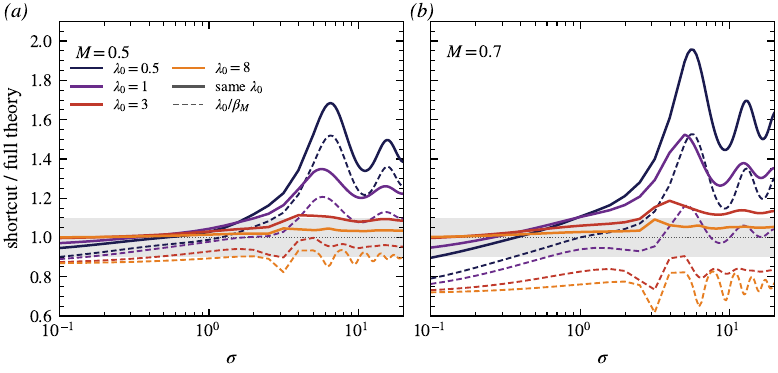}
\caption{Can existing theories be combined? The surviving fraction
$|L_{\rm porous}/L_{\rm rigid}|$ estimated from an incompressible porous theory
and a compressible rigid one, divided by the full compressible result, for
resistive layers at ($a$) $M = 0.5$ and ($b$) $M = 0.7$. Solid, the
incompressible porous load with the same $\lambda_0$; dashed, with
$\lambda_0/\beta_M$. Values above unity understate the load reduction, values
below overstate it; the shaded band is $\pm 10\,\%$.}
\label{fig:hybrid}
\end{figure}

\paragraph{Can existing theories be combined?} Because the permeable load
depends on Mach number less than the rigid one, it is tempting to form the
surviving fraction from two theories that already exist: the treated load from
an incompressible porous theory, which is the $k_e \to 0$ limit of
\eqref{eq:sie} solved by \citet{baddoo2021}, and the reference load from a
compressible rigid one. Figure~\ref{fig:hybrid} tests that shortcut against the
full solution. The rigid load is then exact, so what is plotted is the error of
the porous numerator, in two versions: the incompressible theory used as it
stands, with the same $\lambda_0$ (solid), and with the permeability parameter
corrected by hand to $\lambda_0/\beta_M$ (dashed). At low frequency, where the
chord is compact, the first version is within about $20\,\%$. Over the
non-compact range $\sigma = 5$--$20$ it overestimates the surviving fraction,
and so understates the benefit, by factors of $1.45$--$1.96$ at
$\lambda_0 = 0.5$ and $1.26$--$1.52$ at $\lambda_0 = 1$ when $M = 0.7$
($1.34$--$1.68$ and $1.20$--$1.35$ at $M = 0.5$), by $12$--$16\,\%$ at
$\lambda_0 = 3$ and by at most $6\,\%$ at $\lambda_0 = 8$. Correcting $\lambda$
reduces the error for a weakly permeable surface but overcorrects a strongly
permeable one, whose surviving fraction it underestimates by up to $35\,\%$ at
$\lambda_0 = 8$ and $M = 0.7$. It applies the boundary-condition channel of
figure~\ref{fig:searsmach}($d$) without the kernel channel that offsets it, and
in the seepage limit it returns $\beta_M$ times the true load. The two versions
err in opposite directions, and which of them is closer to the full solution
depends on $\lambda_0$, $M$ and $\sigma$, so the better one cannot be identified
without computing the compressible solution itself.

When the shortcut is adequate follows from the high-frequency limit \eqref{eq:hflimit}.
For heave at high frequency the incompressible porous load exceeds the
compressible one by exactly $(\lambda_0 + M)/\lambda_0$, so the shortcut is
within $10\,\%$ only if $\lambda_0 \gtrsim 10M$, that is $\lambda_0 \gtrsim
5$--$7$ at $M = 0.5$--$0.7$; in steady flow the error is of second order in
$M/\lambda_0$, by \eqref{eq:CLM}. Only for a strongly permeable surface,
therefore, can compressibility be left to the rigid reference. For weakly and
moderately permeable surfaces the compressible porous theory is needed for the
load itself, and for every permeability it is needed for what a ratio of lift
magnitudes does not contain: the phase, the chordwise loading and the
Mach-dependent edge exponent of \S\ref{sec:loading}, and the indicial responses
of \S\ref{sec:indicialres}.

\subsection{Low Mach number, high frequency}
\label{sec:lowmach}

\begin{figure}
\centering
\includegraphics[width=\textwidth]{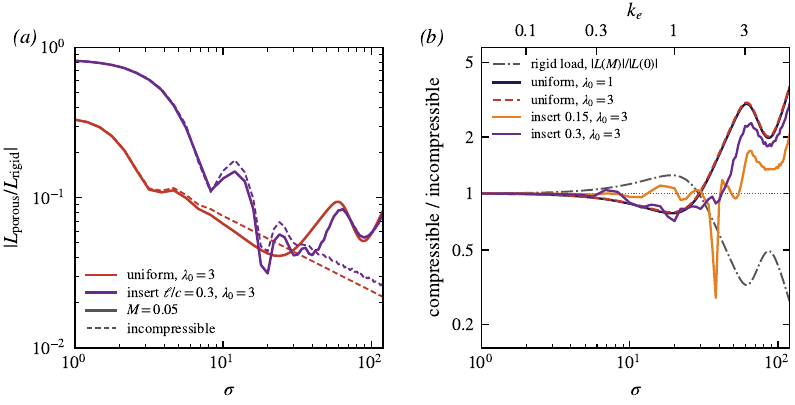}
\caption{Acoustic non-compactness at low Mach number: resistive layers at
$M = 0.05$, compared with the incompressible solution at the same $U$, $\omega$
and material (the same solver at $k_e = 10^{-3}$). ($a$) The surviving fraction
$|L_{\rm porous}/L_{\rm rigid}|$ for a uniformly permeable plate and a
leading-edge insert over $30\,\%$ of the chord, $\lambda_0 = 3$; solid
$M = 0.05$, dashed incompressible. ($b$) The ratio of the compressible to the
incompressible surviving fraction for uniform plates and for inserts over
$15\,\%$ and $30\,\%$ of the chord, and the ratio of the rigid loads
(dash-dot); the upper axis gives $k_e$.}
\label{fig:lowmach}
\end{figure}

The comparisons so far vary the Mach number at reduced frequencies of up to
$20$. The compressible parameter of the kernel is $k_e = M\sigma/\beta_M^2$
(\S\ref{sec:similarity}), so a chord is acoustically non-compact at low Mach
number too once $\sigma$ approaches $1/M$. Figure~\ref{fig:lowmach} takes
$M = 0.05$, at which $\lambda = \lambda_0/\beta_M$ exceeds $\lambda_0$ by
$0.1\,\%$ and the phase $Mk_e$ of \eqref{eq:forcing} is at most $0.3$, so that
the boundary condition and the forcing are effectively incompressible, and
extends the gust response to $\sigma = 120$, where $k_e = 6$ and the chord spans
about two acoustic wavelengths. What separates each pair of curves is then the
acoustic non-compactness of the chord alone.

\paragraph{The rigid load.} Compressibility raises the rigid load by
$12\,\%$ at $\sigma = 10$ and by $25\,\%$ at $\sigma = 20$, and then collapses
it, to $0.61$ of its incompressible value at $\sigma = 40$ and $0.26$ at
$\sigma = 120$ (panel $b$, dash-dot): the low-Mach counterpart of the behaviour
of \S\ref{sec:searsmach}, and the regime of the high-frequency theory of
\citet{amiet1976}.

\paragraph{A uniformly permeable plate.} Its load is set by the material, as in
\S\ref{sec:searsmach}, and the seepage limit \eqref{eq:seep} contains no $k_e$:
the compressible and incompressible loads agree to $3\,\%$ over the whole range
at $\lambda_0 = 1$ and to $1\,\%$ at $\lambda_0 = 3$. The surviving fraction
therefore inherits the non-compactness of the rigid reference entirely. The
compressible fraction is $0.88$--$0.89$ times the incompressible one at $\sigma = 10$
and $0.78$ times it at $\sigma = 20$, and then $1.6$, $3.0$ and $3.7$ times it
at $\sigma = 40$, $60$ and $120$, for both permeabilities.

\paragraph{A leading-edge insert.} Here the porous load is itself non-compact.
The leading edge, the junction and the rigid aft section carry loads that act on
one another with the retarded phase $k_e|\xi - s|$, and over $\sigma =
20$--$120$ the lift of the insert departs from its incompressible value by
factors between $0.55$ and $1.02$ for $\ell/c = 0.3$, and between $0.18$ and
$1.33$ for $\ell/c = 0.15$, whose load nearly cancels near $\sigma = 38$. The
surviving fraction follows neither the incompressible curve nor the rigid
reference. For the $30\,\%$ insert it is $0.71$ of the incompressible value at
$\sigma = 20$ and $1.8$--$3.0$ times it over $\sigma = 60$--$120$; for the
$15\,\%$ insert it is $1.3$--$2.1$ times it over $\sigma = 60$--$120$, and
$0.28$ of it at the near-cancellation.

Non-compactness therefore makes compressibility a low-speed effect as well. At
$M = 0.05$ the surviving fraction departs from the incompressible prediction by
more than $10\,\%$ from $\sigma \approx 10$--$16$, that is $k_e \approx
0.5$--$0.8$, and by factors of $1.3$ to $3.8$ over $\sigma = 60$--$120$ in every
configuration computed. An incompressible porous theory describes the loading
only where $M\sigma \ll 1$, whatever the speed.

\subsection{The porous Theodorsen function}
\label{sec:theo}

\begin{figure}
\centering
\includegraphics[width=\textwidth]{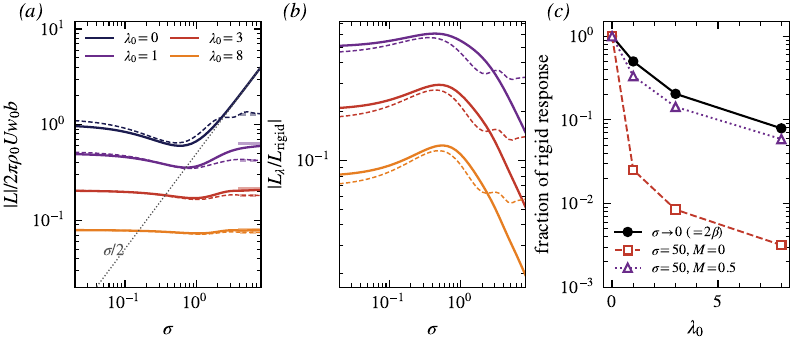}
\caption{Heave response of a permeable plate, $\lambda_0$ independent of
frequency; solid $M = 0$, dashed $M = 0.5$. ($a$) $|L|$ referred to the
incompressible quasi-steady lift of the rigid plate; the rigid incompressible
curve follows $\sigma/2$ at high frequency, the rigid compressible one
saturates at the piston value, and the permeable ones tend to the
high-frequency limit $2/(\pi(\lambda_0 + M))$ of \eqref{eq:hflimit}, which at
$M = 0$ is the seepage limit $2/(\pi\lambda_0)$, marked by the short bars (solid
$M = 0$, broken $M = 0.5$). ($b$) The surviving fraction
$|L_{\rm porous}/L_{\rm rigid}|$: porosity acts as a low-pass filter on the
load. ($c$) The
two ends of the frequency range against $\lambda_0$: the quasi-steady ratio is
the steady closed form $2\beta$, while at $\sigma = 50$ the surviving fraction
is one to two orders of magnitude smaller at $M = 0$, and only $14$--$27\,\%$
smaller at $M = 0.5$, where compressibility has already capped the rigid
response.}
\label{fig:theo}
\end{figure}

For heave the physical upwash is the constant $w = 1$ of \eqref{eq:forcing}, and the rigid
incompressible response is Theodorsen's $\mathcal{C}(\sigma) - \I\sigma/2$,
which the solver returns to $1.4\times10^{-4}$ (table~\ref{tab:verify}). The
porous responses are shown in figure~\ref{fig:theo}. As in \S\ref{sec:searsmach},
the surface is a resistive layer, with $\lambda_0$ real and independent of
frequency. For an inertive perforate $\lambda_0 \propto \I/\sigma$ vanishes at
high frequency, the surface then behaves as a rigid one, and the high-frequency
end of these curves is not reached (\S\ref{sec:inertive}).

\paragraph{Porosity removes the apparent mass almost entirely.} The rigid
response grows without bound, $|L| \to \sigma/2$, because accelerating a rigid
plate accelerates the fluid attached to it. A permeable plate cannot sustain the
pressure jump that would require: the seepage term $\lambda\Pi$ grows with the
loading, and the response tends to a \emph{constant} instead. At $M = 0$ and
$\sigma = 50$ the rigid value is $25.0$ and the porous ones are $0.630$,
$0.212$ and $0.0795$ for $\lambda_0 = 1$, $3$ and $8$: fractions of
$2.5\times10^{-2}$, $8.5\times10^{-3}$ and $3.2\times10^{-3}$. The constants
are the seepage limit \eqref{eq:seep}, $2/(\pi\lambda_0) = 0.637$, $0.212$ and
$0.0796$, to within $1\,\%$: at high frequency the load of a resistive plate is
set by the material, which has no inertia to add, and not by the fluid that a
rigid plate would have to accelerate. A layer with inertia keeps part of its
apparent mass (\S\ref{sec:inertive}). At the other end
of the frequency range the quasi-steady limit is the steady closed form
\eqref{eq:CL}, and the solver returns it to four figures: at $\sigma = 10^{-4}$, $0.5000$, $0.2048$ and $0.0792$ against $2\beta = 0.5000$,
$0.2048$ and $0.0792$; at $M = 0.5$ the ratio to the rigid plate at the same
Mach number is the same closed form evaluated with $\lambda_0/\beta_M$, to the
same accuracy. Between those two
ends lies the whole content of panel ($c$): \emph{for a resistive layer in incompressible flow porosity is far more
effective against the non-circulatory load than against the circulatory one}: at
$\lambda_0 = 3$ the apparent mass falls to $0.3\,\%$ of its rigid value while
the circulation keeps $20\,\%$. For flutter this is the useful direction, since apparent mass is what couples
plunge to pitch inertia, and it is the opposite of what the steady lift ratio
alone would suggest.

\paragraph{Compressibility caps the rigid response and moves the permeable one
much less.} At $M = 0.5$ the rigid heave response no longer grows like
$\sigma/2$: it saturates at the acoustic (piston) value $2/(\pi M) = 1.27$ of
\S\ref{sec:forcing}, the classical result that at high reduced frequency each
element of the surface radiates as a piston. The permeable responses tend to
the high-frequency limit \eqref{eq:hflimit} instead (panel $a$): at $\sigma = 50$ they
are $0.42$, $0.18$ and $0.075$ for $\lambda_0 = 1$, $3$ and $8$, equal to
$2/(\pi(\lambda_0 + M))$ to four figures, against $0.630$, $0.212$ and $0.0795$
at $M = 0$. The change, $M/(\lambda_0 + M)$, is $33$, $14$ and $6\,\%$, and it is smaller
the more permeable the surface, as in \S\ref{sec:searsmach}. The apparent mass that
porosity would have removed has therefore already been removed by
compressibility, and the surviving fractions at $\sigma = 50$ rise to $0.33$,
$0.14$ and $0.059$, thirteen to nineteen times the incompressible fractions,
so that the benefit is an order of magnitude smaller than the incompressible
calculation reports. This is the Sears result of \S\ref{sec:searsmach} again:
a nearly Mach-independent permeable load measured against a rigid load that
carries the whole of the classical compressibility effect.

\subsection{Indicial responses}
\label{sec:indicialres}

\begin{figure}
\centering
\includegraphics[width=\textwidth]{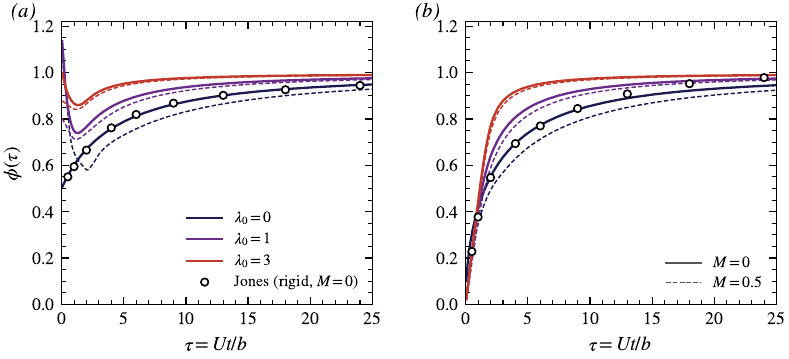}
\caption{Indicial responses, each normalised by its own quasi-steady value.
($a$) The Wagner analogue: the response to a step in incidence, with the
apparent-mass impulse removed as appendix~\ref{app:indicial} describes.
($b$) The K\"ussner analogue: the response to a sharp-edged gust, with $\tau$
measured from the instant the gust front reaches the leading edge. Solid
$M = 0$, dashed $M = 0.5$; symbols, Jones's approximations to the classical
rigid functions at $M = 0$.}
\label{fig:indicial}
\end{figure}

Figure~\ref{fig:indicial} gives the porous analogues of the Wagner and
K\"ussner functions, obtained from \eqref{eq:indicial} as
appendix~\ref{app:indicial} describes. Each is normalised by its \emph{own}
quasi-steady value, so the curves show how fast the load settles, not how large
it is; the size is the steady factor $2\beta$ of \eqref{eq:CL}, and the two
statements should be read together.

\paragraph{The rigid limit.} At $M = 0$ and $\lambda = 0$ the computed Wagner
function agrees with Jones's classical approximation to $0.007$ over
$\tau \le 40$, including the correct initial value $\phi(0^+) = 1/2$, which is not imposed
anywhere and emerges from the high-frequency limit of the
computed transfer function. The K\"ussner analogue agrees to $0.035$ for
$\tau \ge 2$ and departs by up to $0.095$ below $\tau = 1$, where Jones's
two-exponential fit is itself least reliable.

\paragraph{Porosity shortens the transient as well as reducing the load.} A
permeable plate reaches $90\,\%$ of its steady gust response in $\tau = 7.4$
semichords at $\lambda_0 = 1$ and $3.5$ at $\lambda_0 = 3$, against $14.3$ for
a rigid plate: the treatment removes most of the wake memory along with most of
the load. The mechanism is the same one that sets $2\beta$: less circulation is shed, so
the wake exerts less influence on what follows. The consequence is that a permeable section responds to a passing vortex with a
load that is both smaller and shorter, which is not what a quasi-steady
estimate scaled by $2\beta$ would give.

\paragraph{Compressibility slows the response and changes how it starts.} At
$M = 0.5$ the rigid Wagner analogue begins near the piston value, since the
first instant after a step in incidence is the high-frequency limit of
\S\ref{sec:theo}; the computation gives $1.08$ of the steady lift, against
$1.10$ in the limit. It then falls within two semichords to $0.58$ and recovers
more slowly than the incompressible function, reaching $90\,\%$ of its steady
value at $\tau = 18$ rather than $13$. The K\"ussner
analogue is delayed: it has reached $0.10$ of its steady value at
$\tau = 0.25$, against $0.22$ incompressibly, and $90\,\%$ at $\tau = 19.4$
against $14.3$, because the gust front takes a finite time to make itself felt
through a non-compact chord. The permeable cases inherit the delay in weaker
form: at $\lambda_0 = 3$ the K\"ussner rise time lengthens from $3.5$ to $3.9$
semichords. They do not inherit the raised start. A permeable plate has almost
no apparent mass even at $M = 0$, so its Wagner analogue already begins near its
steady value, and compressibility lowers that start, to $0.88$ at
$\lambda_0 = 3$, because the piston-limited high-frequency response is then
below the permeable steady lift. For a gust-encounter or blade--vortex interaction calculation the practical
consequence is that the transient a permeable section produces at flight Mach number is neither the incompressible porous one nor the
compressible rigid one.

\subsection{The steady limit: lift and seepage drag}
\label{sec:steady}

At $\omega = 0$ the same boundary condition is a complete steady thin-aerofoil
theory, and it closes in terms of the single exponent \eqref{eq:beta}. For a
uniformly porous plate at incidence $\alpha_i$ the Riemann--Hilbert solution is
a single power law and (appendix~\ref{app:steady})
\begin{align}
\frac{C_L}{C_{L,\rm rigid}} &= 2\beta = 1 - \frac{2}{\pi}\arctan\lambda,
\label{eq:CL}\\
C_{d,\rm seep} &= \alpha_i\,C_L .
\label{eq:Cd}
\end{align}
Equation~\eqref{eq:Cd} is written for a flat plate at incidence, where the
physical upwash is the constant $w = \alpha_i$ and $\Wup = -\alpha_i$. For a
general mean line the streamwise force is the loading worked against that
upwash, which in the normalisation of appendix~\ref{app:steady}, where $C_L =
-2\int\Pi\,\D\xi$, is
\begin{equation}
C_{d,\rm seep} = 2\int_{-1}^{1}\Pi(\xi)\,\Wup(\xi)\,\D\xi ,
\label{eq:Cdgen}
\end{equation}
of which \eqref{eq:Cd} is the constant-$\Wup$ case. The distinction matters
wherever camber is used to buy the penalty back: a cambered section at its
zero-lift incidence carries loading that integrates to zero but is not zero
pointwise, and since the dissipation is quadratic in the loading, the
cancellation that removes the lift leaves the drag untouched. Seepage drag
vanishes with zero \emph{loading}, not with zero \emph{lift}.

\begin{figure}
\centering
\includegraphics[width=\textwidth]{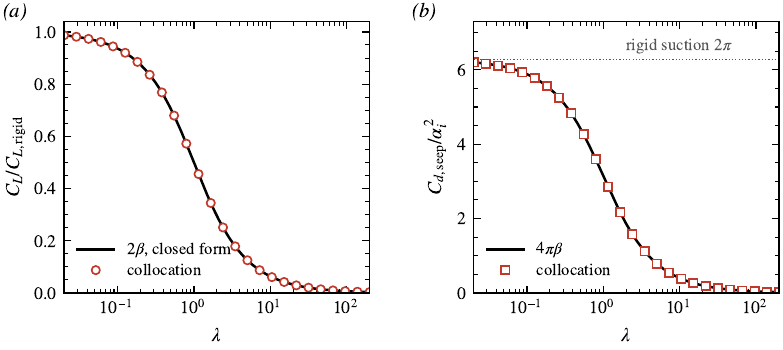}
\caption{The steady limit, uniformly permeable plate. ($a$) The lift ratio:
the closed form $2\beta$ of \eqref{eq:CL} (line) against the collocation solver
(symbols), over four decades of $\lambda$, from $0.02$ to $200$. ($b$) The seepage drag per unit
incidence squared, $4\pi\beta$ from \eqref{eq:Cd}; the dotted line is the
leading-edge suction $2\pi$ that a rigid edge would have recovered. Solver and
closed forms agree to $4\times10^{-4}$.}
\label{fig:steady}
\end{figure}

Three statements follow, and all three are consequences of the edge exponent
rather than of the material as such. First, the lift ratio is exactly twice the
edge exponent, so one number sets the loss of lift and the weakening of the
singularity together (figure~\ref{fig:steady}): half the lift is gone at
$\lambda = 1$, where $\beta = 1/4$. Second, a rigid plate at incidence carries a
streamwise pressure-force component $\alpha_i C_{L,\rm rigid}$ that is cancelled
exactly by leading-edge suction, and that suction exists only because the
loading is singular as $(1+\xi)^{-1/2}$; a permeable edge carries
$(1+\xi)^{-\beta}$ with $\beta < 1/2$, so $(1+\xi)\Pi^2 \to 0$ at the nose,
there is no suction, and \eqref{eq:Cd} says the streamwise component is
dissipated entirely by seepage, at $2\beta$ times the suction $2\pi\alpha_i^2$
that the corresponding rigid edge would have developed. Third, the drag integrand
behaves as $(1+\xi)^{-2\beta}$, integrable if and only if $\beta < 1/2$, so the
porosity-dependent exponent is what keeps the energy budget finite. The drag
therefore falls as $\lambda$ increases (figure~\ref{fig:steady}$b$): the seepage
velocity grows, but the pressure jump that drives it falls faster, and the
largest penalty, close to the full suction $2\pi\alpha_i^2$, is paid by a
slightly permeable edge.

\paragraph{Mach number.} In compressible flow $C_{L,\rm rigid} = 2\pi\alpha_i/\beta_M$,
and \eqref{eq:CL} holds with the local value of $\lambda$. When the compressible
and incompressible theories are compared at the same flow speed and material,
$\lambda = \lambda_0/\beta_M$ by \eqref{eq:lamM} for any material, and the lift
therefore varies with Mach number as
\begin{equation}
\frac{C_L(M)}{C_L(0)} = \frac{1 - (2/\pi)\arctan(\lambda_0/\beta_M)}
{\beta_M\,\big[1 - (2/\pi)\arctan\lambda_0\big]} ,
\label{eq:CLM}
\end{equation}
which is the Prandtl--Glauert factor $1/\beta_M$ for a rigid plate and tends to
unity as $\lambda_0 \to \infty$, where $C_L \to 4\alpha_i/\lambda_0$. In that
limit the Prandtl--Glauert rise of the circulation is cancelled exactly by the
rise of $\lambda$, and the lift is the seepage load \eqref{eq:seep} of a surface
that passes the incidence upwash. At $M = 0.7$ the steady lift rises by
$10.6$, $1.7$ and $0.25\,\%$ for $\lambda_0 = 1$, $3$ and $8$, against
$40\,\%$ for a rigid plate, and the unsteady solver reproduces \eqref{eq:CLM}
to three figures in its quasi-steady limit. For a resistive layer the limit is
the Darcy drop $\Delta p = Rw$, which contains no Mach number.

\begin{figure}
\centering
\includegraphics[width=\textwidth]{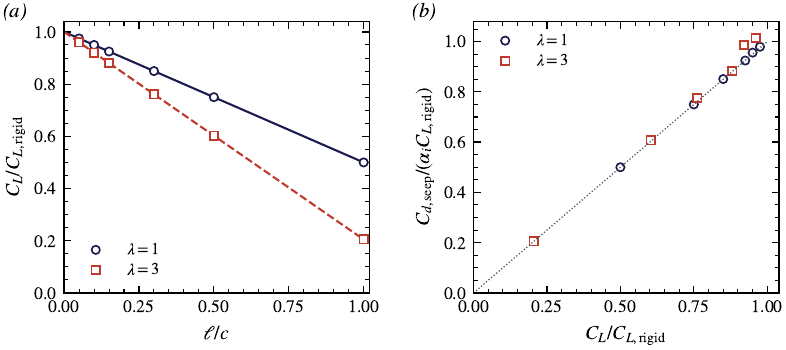}
\caption{Steady penalty of a leading-edge insert. ($a$) Lift against treated
length: symbols, the collocation solver with junction modes; lines, the linear
law \eqref{eq:CLinsert}. ($b$) Seepage drag against lift, both per unit
incidence; the identity $C_{d,\rm seep} = \alpha_i C_L$ is the dotted
diagonal.}
\label{fig:insert}
\end{figure}

For a permeable insert of length $\ell$ at the leading edge the solver returns
something simpler than the local analysis would suggest
(figure~\ref{fig:insert}). Over $\ell/c = 0.05$--$1$ and $\lambda = 0.5$--$10$
the computed lift satisfies
\begin{equation}
\frac{C_L}{C_{L,\rm rigid}} = 1 - \frac{\ell}{c}\,(1 - 2\beta)
\label{eq:CLinsert}
\end{equation}
to $10^{-5}$. It is a linear interpolation between the rigid plate and the
uniformly permeable closed form \eqref{eq:CL}, with no dependence on where
between the two the junction sits beyond its position. The lift penalty is
therefore proportional to the treated chord fraction, which is the design
statement one would want and is not obvious from a loading distribution that
carries two algebraic singularities. The law is exact: for a constant upwash the
steady Riemann--Hilbert problem with a piecewise-constant $\lambda$ has a
closed-form solution whose lift is linear in the junction position
(appendix~\ref{app:steady}), and the solver reproduces it to $10^{-5}$.

The seepage-drag identity survives the junction. With the forcing normalised to
unit incidence, $C_{d,\rm seep} = \alpha_i C_L$ becomes the statement that the
two computed coefficients are equal, and they are: to four figures for a uniform
plate, and to within $7\,\%$ for a partial insert, the largest departures
being for short inserts at $\lambda = 3$, where the drag integrand is sampled
right up to the junction singularity and converges most slowly.

\subsection{A layer with inertia}
\label{sec:inertive}

\begin{figure}
\centering
\includegraphics[width=\textwidth]{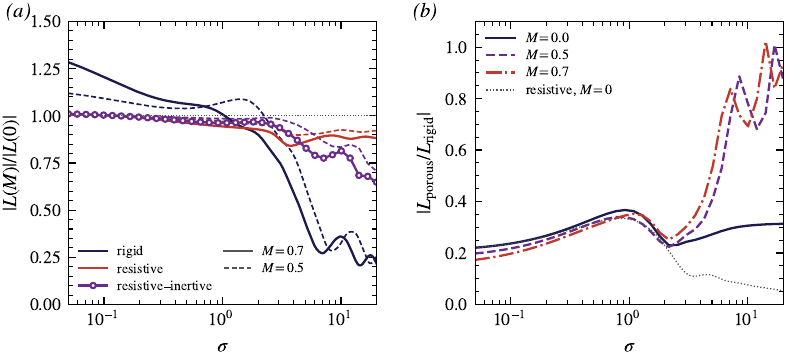}
\caption{Gust response of a resistive--inertive layer,
$\lambda_0 = \lambda_R/(1 - \I\sigma/\sigma_m)$ with $\lambda_R = 3$ and
$\sigma_m = 2$, against the resistive layer with $\lambda_0 = 3$ and the rigid
plate. ($a$) The Mach factor $|L(M)|/|L(0)|$ of each surface at $M = 0.7$
(solid) and $0.5$ (dashed). ($b$) The fraction of the rigid load at the same Mach
number that survives, for the resistive--inertive layer at $M = 0$, $0.5$ and
$0.7$; dotted, the resistive layer at $M = 0$.}
\label{fig:inertive}
\end{figure}

A real perforate or porous layer has inertia as well as resistance, since the
fluid in the pores must be accelerated, and the material law of
\citet{baddoo2021} carries an effective density for that reason. Writing $Z_d =
R - \I\omega m_a$ in \eqref{eq:lamM} gives
\begin{equation}
\lambda_0(\sigma) = \frac{\lambda_R}{1 - \I\sigma/\sigma_m},
\qquad \lambda_R = \frac{2\rho_0U}{R}, \qquad \sigma_m = \frac{Rb}{Um_a},
\label{eq:inertive}
\end{equation}
resistive below $\sigma_m$ and inertive above it, where $|\lambda_0|$ falls as
$1/\sigma$. Figure~\ref{fig:inertive} shows the gust response for $\lambda_R =
3$ and $\sigma_m = 2$: the resistive layer of \S\ref{sec:searsmach} below
$\sigma \approx 2$, and above it a surface whose effective permeability falls to
$|\lambda_0| = 1.1$ at $\sigma = 5$ and $0.42$ at $\sigma = 14$.

\paragraph{Permeability still controls the sensitivity to compressibility.} The
Mach factor of panel ($a$) follows the resistive layer where the two have the
same $\lambda_0$, lying between $0.96$ and $1.01$ at $M = 0.7$ for $\sigma \le
2$, and moves towards the rigid plate as $|\lambda_0|$ falls: at $M = 0.7$ it is
$0.83$ at $\sigma = 5$ and $0.65$ at the top of the range, against
$0.86$--$0.90$ for the resistive layer and $0.2$--$0.4$ for a rigid plate. The
same statement covers both materials: the sensitivity of the load to
compressibility is set by how permeable the surface is at the frequency in
question. It also defeats the shortcut of \S\ref{sec:searsmach}: at
$\sigma = 20$ and $M = 0.7$ the incompressible load of this layer is $1.5$ times
the compressible one, because the surface has become weakly permeable where the
chord is non-compact.

\paragraph{Inertia removes most of the high-frequency benefit.} Incompressibly
the fraction of the rigid gust load that survives levels off at $0.27$--$0.31$
for $\sigma \ge 5$, where the resistive layer falls to $0.05$ (panel $b$). At $M
= 0.5$ and $0.7$ it rises to between $0.4$ and $1.0$, $1.5$ to $3.3$ times the
incompressible value, and near the interference minima of the rigid response the
treatment achieves nothing. The heave response makes the same point: at $M = 0$
this layer keeps $17\,\%$ of the rigid plate's apparent mass at $\sigma =
20$--$50$, against $0.85\,\%$ at $\sigma = 50$ for the resistive layer with
$\lambda_0 = 3$, because its admittance falls with frequency and the
high-frequency load is again supported by the flow. At $M = 0.5$ it keeps
$82$--$96\,\%$ of the rigid piston load over the same range, against
$14\,\%$ for the resistive layer. The removal of apparent mass found in \S\ref{sec:theo} is
a property of resistive layers.

\subsection{Limits of validity}
\label{sec:limits}

The results of \S\S\ref{sec:searsmach}--\ref{sec:steady} are for resistive
layers, and \S\ref{sec:inertive} shows how inertia changes them; the formulation
itself admits any admittance, complex and varying along the chord. The
formulation is linear, subsonic and thin-aerofoil, so it inherits the limits
of the Prandtl--Glauert transformation and the Possio kernel: a thin section,
attached flow, small incidence, and no transonic region, which in practice means
$M \lesssim 0.7$. Nothing in the edge analysis degrades as $M$ rises, since the
exponents are independent of $k_e$, and the solver remains cheap there: at
$M = 0.6$ and $\kb = 20$, so $k_e = 12$, the lift is converged to six figures at
$N = 140$. Two further conditions belong to the material rather than to the
aerodynamics. The homogenised admittance requires many apertures per
hydrodynamic wavelength, which is a statement about the perforate pitch and the
frequency, and becomes demanding at the reduced frequencies of
\S\ref{sec:lowmach}, where $\sigma = 120$ puts the hydrodynamic wavelength at
$2\pi b/\sigma \approx 0.05b$; and for a bulk medium the pore Reynolds number must be of order unity
or below, since the leading correction to Darcy's law is Forchheimer's
\citep{forchheimer1901}, in which the resistance grows with the seepage velocity
and hence with the gust amplitude, an amplitude dependence that a linear
frequency-domain theory cannot carry. The seepage limit \eqref{eq:seep} makes that
condition more pressing rather than less: a surface that passes the imposed
upwash carries seepage velocities of the order of the gust velocity itself.

\section{Conclusions}
\label{sec:concl}

A linear theory has been developed for the unsteady loading of finite-chord
porous aerofoils in compressible subsonic flow. It combines Possio's integral
operator, written with its Cauchy singularity separated from a bounded
remainder, with a convective permeable boundary condition, allows
chordwise-varying admittance, and retains the shed wake and the unsteady Kutta
condition; the material enters through the permeability parameter $\lambda =
2\Ad M/\beta_M$ alone. Because the non-Cauchy part of the operator is bounded,
the loading exponents at the aerofoil edges and at admittance discontinuities
retain their incompressible functional form when expressed in terms of the local
permeability parameter, and are independent of the reduced frequency. The
weighted-Jacobi collocation method therefore carries over, with the
trailing-edge exponent carried as a parameter and junction modes added where the
admittance jumps. The solution reproduces
the published incompressible code of \citet{baddoo2021} to $0.09\,\%$ and agrees
with an independent compressible formulation in physical variables to $10^{-3}$
or better.

The harmonic gust and heave responses, together with their indicial
counterparts, show that permeability controls the sensitivity of the unsteady
loading to compressibility. The results are for resistive layers, compared at
the same flow speed, frequency and material, so that $\lambda =
\lambda_0/\beta_M$ for any material. The pressure jump across a permeable surface is set
partly by the surrounding flow and partly by the material impedance, and
compressibility acts on the first. As permeability increases the material
governs an increasing share of the load, and the Mach-number dependence
associated with the surrounding flow weakens: at $M = 0.7$ and $\sigma = 5$ the
gust load is $0.43$ of its incompressible value for a rigid plate, $0.53$ at
$\lambda_0 = 0.5$, $0.66$ at $\lambda_0 = 1$, $0.86$ at $\lambda_0 = 3$ and
$0.95$ at $\lambda_0 = 8$. The weakening is gradual rather than complete: a
weakly permeable surface keeps much of the Mach dependence of a rigid plate, and
only a strongly permeable one loses it. The Mach number reaches the load through
two channels that act in opposite directions. The boundary condition, through which a given
material appears more permeable by $1/\beta_M$, lowers the load by a fifth to a
quarter at $M = 0.7$, nearly independently of frequency; the integral operator,
which carries the Prandtl--Glauert rise, raises it at low frequency and, unlike
its effect on a rigid plate, does not collapse it at high frequency. In the time
domain a permeable section responds faster as well as more weakly, reaching
$90\,\%$ of its steady gust response in $3.5$ semichords at $\lambda_0 = 3$
against $14.3$ for a rigid one, and compressibility lengthens these rise times
to $3.9$ and $19.4$ semichords at $M = 0.5$.

Closed-form steady and high-frequency limits describe this transition for
resistive surfaces. In steady
flow the lift varies with Mach number as \eqref{eq:CLM}, between the
Prandtl--Glauert factor of a rigid plate and unity, and tends to $C_L =
4\alpha_i/\lambda_0$ as permeability increases, with a residual Mach dependence
of second order in $M/\lambda_0$. At high frequency the heave load tends to
$2/(\pi(\lambda_0 + M))$ of the incompressible rigid quasi-steady lift: the
material and acoustic (piston) resistances of the two faces act in parallel,
since both are driven by the same pressure jump, giving the piston value for a rigid plate and the pressure drop set by the
material impedance, $\rho_0c_0w/\Ad$, for a strongly permeable one, with a
residual Mach dependence of first order in $M/\lambda_0$. Together the two
limits state when compressibility can be ignored for the porous load: only when
$\lambda_0 \gg M$. The same limit explains why, for a resistive layer in incompressible flow,
porosity removes the apparent mass far more effectively than the circulation:
the high-frequency load is set by a material that has no inertia to add. A layer
with inertia as well as resistance becomes less permeable as the frequency
rises; its Mach sensitivity then moves towards that of a rigid plate, it keeps
part of its apparent mass, $17\,\%$ in the case computed, and it loses most of
its high-frequency benefit.

Although the porous load becomes less sensitive to Mach number, its ratio to the
impermeable load remains strongly affected by compressibility, because the
impermeable load follows the classical behaviour: a Prandtl--Glauert rise at low
frequency and a collapse once the chord is acoustically non-compact, to a
quarter of its incompressible value at $\sigma = 20$ and $M = 0.5$.
Incompressible theory overestimates the ratio at low frequency, most of all in
steady flow, where the compressible ratio is the incompressible one times
$2\beta(\lambda_0/\beta_M)/2\beta(\lambda_0)$, $0.87$ at $M = 0.5$ and $0.73$
at $M = 0.7$ for $\lambda_0 = 3$; it underestimates the ratio once the chord is
non-compact. For the gust cases examined at $M = 0.5$--$0.7$ and $\sigma =
5$--$20$ it underestimates the ratio by factors of $1.4$--$4.3$ ($7.6\,\%$
against $21.6\,\%$ at $\sigma = 10$ and $M = 0.5$), thereby overstating the load reduction achieved by the treatment;
for the layer with inertia the factor reaches $3.3$, and near the interference
minima of the rigid response the treatment achieves nothing. The heave response shows the same
structure: the unbounded apparent mass of a rigid plate is capped by
compressibility at its piston value, and the fraction of the rigid
high-frequency load that a permeable plate retains at $\lambda_0 = 3$ rises from
under $1\,\%$ incompressibly to $14\,\%$ at $M = 0.5$.

It follows that the existing theories cannot simply be combined. Taking the
porous load from incompressible theory and the rigid load from compressible
theory overestimates the surviving fraction, over $\sigma = 5$--$20$ at
$M = 0.7$, by factors of up to $1.96$ at $\lambda_0 = 0.5$ and $1.52$ at
$\lambda_0 = 1$; correcting the permeability parameter by $1/\beta_M$
overcorrects a strongly permeable surface, underestimating its surviving
fraction by up to $35\,\%$ at $\lambda_0 = 8$; and for the layer with inertia,
whose permeability falls with frequency, the shortcut is $1.5$ times too high at
$\sigma = 20$. The high-frequency limit sets the condition for the shortcut,
$\lambda_0 \gtrsim 10M$ for $10\,\%$ accuracy, and even where that holds a ratio
of lift magnitudes supplies neither the phase, nor the chordwise loading and the
Mach-dependent edge exponent on which scattering depends, nor the indicial
responses. A theory that carries permeability and compressibility together is
therefore required for the load of a weakly or moderately permeable aerofoil,
and for everything beyond the lift magnitude at any permeability.

Nor is compressibility only a high-speed effect. The kernel carries the acoustic
compactness $k_e = M\sigma/\beta_M^2$ separately from the Mach number of the
boundary condition, and at $M = 0.05$ the surviving fraction departs from its
incompressible value by more than $10\,\%$ from $\sigma \approx 10$, and by
factors of up to $3.8$ at $\sigma = 120$, where the chord spans about two
acoustic wavelengths. For a uniformly permeable plate this comes from the rigid
reference alone; for a leading-edge insert the porous load is itself
non-compact, the leading edge, the junction and the aft section acting on one
another with a retarded phase. An incompressible porous theory therefore
describes the loading only where $M\sigma \ll 1$, at any speed.

In the steady limit the lift ratio is $C_L/C_{L,\rm rigid} = 2\beta$ and the
seepage drag is $C_{d,\rm seep} = \alpha_i C_L$, so the loss of lift and the
weakening of the edge singularity are one number, and a permeable edge develops
no leading-edge suction. For a leading-edge insert the lift penalty is, to
$10^{-5}$, linear in the treated fraction of the chord, $C_L/C_{L,\rm rigid} = 1
- (\ell/c)(1-2\beta)$.

The formulation is thus a unified linear theory of the unsteady loading of a
thin aerofoil, rigid or permeable, in incompressible or subsonic compressible
flow, with the classical and incompressible porous results as its limits. It
provides these loads at the cost of a rigid thin-aerofoil calculation, for
assessing porous aerofoils and for supplying aerodynamic inputs to aeroelastic
and aeroacoustic models.

\backsec{Acknowledgements}{%
The author used Claude (Anthropic), a large language model, in the versions
Claude Opus~5, Claude Opus~5.5 and Claude Fable~5.1, accessed through the
Claude desktop application, in the course of this work in September 2026. It
was used to check the analytical derivations, including the signs and
conventions of the governing equations, to write and run, under the author's
direction, the numerical scripts from which the results and figures were
produced, to port and run the published code of \citet{baddoo2021} for the
comparison of \S\ref{sec:verify}, to check the bibliography against publisher
records, and to assist in drafting and editing the text. Every derivation,
numerical result and statement in the paper was examined and verified by the
author, who takes full responsibility for its content. The model is not an
author of this work.}

\backsec{Funding}{This research received no specific grant from any funding
agency, commercial or not-for-profit sectors.}

\backsec{Declaration of interests}{The author reports no conflict of interest.}

\backsec{Data availability statement}{The data and scripts that support the
findings of this study are available from the corresponding author upon
reasonable request.}

\backsec{Author ORCID}{S. Lee, \url{https://orcid.org/0000-0001-8081-2268}.}

\appendix
\renewcommand{\thesection}{\Alph{section}}

\section{The convective permeable boundary condition}
\label{app:bc}

Let $\varphi$ be the disturbance velocity potential and $[\,\cdot\,]$ the jump
across the sheet. Three ingredients are combined. \emph{(i) Linearised momentum}
across the sheet gives
$\Delta p = -\rho_0(\partial_t + U\partial_{x_1})[\varphi]$. \emph{(ii) Mass
conservation} requires the total normal velocity to be continuous and equal to
the seepage, $w + \partial_{x_2}\varphi = v_s$ on both faces. \emph{(iii) The
material law} \eqref{eq:vs}, $v_s = -\Ad\Delta p/(\rho_0c_0)$. Eliminating $\Delta p$
between (i) and (iii) and substituting into (ii) gives \eqref{eq:bcconv}. The
order matters: (ii) is the only ingredient carrying boundary information, and
the seepage enters as the \emph{normal velocity}, not as a second statement
about $\Delta p$.

To obtain the coefficient of $\Pi$ in \eqref{eq:sie}, write
$\Delta p = C\rho_0Uw_0\Pi\,\E^{-\I Mk_e\xi}$ for whatever constant $C$ the
normalisation of $\Pi$ implies, the exponential being the Prandtl--Glauert
factor of appendix~\ref{app:kernel}. Then (iii) contributes
$-\Ad CMw_0\Pi\,\E^{-\I Mk_e\xi}$ to the physical normal velocity, and so
$-\Ad CMw_0\Pi$ to the transformed one, which appears with a positive sign on
the left of \eqref{eq:sie}, while the
flat-plate inversion of (ii) contributes
$(\beta_MC/2\pi)\int\Pi(s)/(\xi-s)\,\D s$ after the Prandtl--Glauert
transformation. Fixing the Cauchy coefficient at $1/\pi$ requires
$C = 2/\beta_M$, whence $\lambda = \Ad CM = 2\Ad M/\beta_M$ as claimed. Because
$\lambda$ is the ratio of the seepage coefficient to the Cauchy coefficient it
is independent of how $\Pi$ is normalised. Holding $\mu = -\I k_0\Ad$ fixed as
$M \to 0$, in which case $\Ad = \I\mu/(Mk_1)$ diverges, keeps $\lambda =
2\I\mu/(\beta_Mk_1)$ finite. The resulting condition still contains the
convective derivative, which the linearised Bernoulli relation carries in
incompressible flow as well; the static porous half-plane condition of
\citet{jaworski2013} and \citet{ayton2021}, $\partial_{x_2}\varphi \propto
[\varphi]$, is recovered only when that derivative is dropped, as at $U = 0$.

\section{From the convected wave equation to \eqref{eq:sie}}
\label{app:kernel}

The operator that follows is classical: it is the subsonic unsteady kernel of
\citet{possio1938}, and no novelty is claimed for it. It is set down here in a
particular \emph{form}, with the Cauchy singularity separated from a remainder
that is shown to be bounded, because that separation is what the
edge analysis of \S\ref{sec:edge} and the quadrature of \S\ref{sec:quad} both
act on, and the classical representations, which carry the singularity inside
the special functions, do not supply it.

\paragraph{Prandtl--Glauert.} The disturbance potential satisfies
$c_0^{-2}(\partial_t + U\partial_{x_1})^2\varphi = \nabla^2\varphi$, which for
$\E^{-\I\omega t}$ is
$\beta_M^2\varphi_{x_1x_1} + \varphi_{x_2x_2} + 2\I Mk_0\varphi_{x_1}
+ k_0^2\varphi = 0$. Writing $\varphi = \Psi\E^{\I\mu x_1}$, the coefficient of
$\Psi_{x_1}$ vanishes for $\mu = -Mk_0/\beta_M^2$, leaving
$\beta_M^2\Psi_{x_1x_1} + \Psi_{x_2x_2} + (k_0/\beta_M)^2\Psi = 0$. The stretch
$X = x_1/\beta_M$ then gives the Helmholtz equation
\begin{equation}
\Psi_{XX} + \Psi_{x_2x_2} + \kappa^2\Psi = 0, \qquad \kappa = k_0/\beta_M ,
\label{eq:helm}
\end{equation}
on a plate of half-length $\tilde b = b/\beta_M$. Normalising $\xi = X/\tilde b$
gives $k_e = \kappa\tilde b = k_0b/\beta_M^2$ as in \eqref{eq:nondim}; the same
bookkeeping on the gust, whose chordwise wavenumber becomes
$k_1 - \mu = k_1/\beta_M^2$, gives $\kb = k_1b/\beta_M^2$ and hence
$k_e = M\kb$. Since $X/\tilde b = x_1/b$, $\xi$ is also the physical chordwise
coordinate, and $\mu x_1 = -Mk_e\xi$: the two potentials are related by
$\varphi = \Psi\,\E^{-\I Mk_e\xi}$. The material derivative transforms as
\[
\Big(\frac{\partial}{\partial t} + U\frac{\partial}{\partial x_1}\Big)\varphi
= \frac{U}{b}\,\E^{-\I Mk_e\xi}\Big(\frac{\partial}{\partial\xi} - \I\kb\Big)\Psi ,
\]
because $\omega - U\mu = \omega/\beta_M^2$. This is why the transformed loading
\eqref{eq:PiFromM} carries $\kb$ rather than $\sigma$, and why the physical
pressure jump is $\Pi\,\E^{-\I Mk_e\xi}$ and a physical upwash enters
\eqref{eq:sie} multiplied by $\E^{\I Mk_e\xi}$ (\S\ref{sec:forcing}).

\paragraph{The dipole sheet.} With $m$ the potential jump and
$\gamma = -\partial m/\partial\xi$ the bound vorticity, the loaded plate is a sheet of normal doublets with potential $\varphi =
-\partial_{x_2}\!\int m\,G\,\D s$, which jumps by $[\varphi] = \varphi(0^+) -
\varphi(0^-) = m$ across the sheet, and the normal velocity it induces on itself
is $w = -\partial^2_{x_2}\!\int m\,G\,\D s$, with $G = (\I/4)H_0^{(1)}(\kappa
R)$ the outgoing Green's function of \eqref{eq:helm}. That integral is hypersingular
as written; \eqref{eq:helm} itself removes the difficulty, since
$G_{x_2x_2} = -G_{XX} - \kappa^2G$ away from the source, so that $w = \int
m\,(G_{XX} + \kappa^2G)\,\D s$. Integrating the first part by parts once, with the boundary terms vanishing
because $m(-1) = 0$ by Kelvin's theorem \eqref{eq:kelvin} and the sheet
continues into the wake at $\xi = 1$, and using $\D H_0^{(1)}/\D z =
-H_1^{(1)}$,
\begin{equation}
w(\xi) = \frac{\I k_e}{4}\!\int_{-1}^{1}\! H_1^{(1)}(z)\,\mathrm{sgn}(\xi-s)\,
\gamma(s)\,\D s
\;+\; \frac{\I k_e^2}{4}\!\int_{-1}^{1}\! H_0^{(1)}(z)\,m(s)\,\D s ,
\qquad z = k_e|\xi-s| ,
\label{eq:possio}
\end{equation}
plus the wake integral. The second term carries $k_e^2$ and has no
incompressible counterpart.

\paragraph{The singular part.} Since $H_1^{(1)}(z)\sim-2\I/(\pi z)$, the first
kernel tends to $1/[2\pi(\xi-s)]$ as $z \to 0$: the classical Glauert kernel,
coefficient and all. Splitting
$H_1^{(1)}(z) = -2\I/(\pi z) + [H_1^{(1)}(z) + 2\I/(\pi z)]$, in which the
bracket is bounded, turns \eqref{eq:possio} into a Cauchy principal value plus
bounded terms.

\paragraph{Assembly.} Adding the seepage contribution (appendix~\ref{app:bc})
and changing the unknown to $\Pi$ through $\Pi = \gamma + \I\kb m$ gives
\eqref{eq:sie}. The change of dependent variable contributes one further term.
On the wake $\Pi = 0$, so that $\gamma = -\I\kb m$ there, and the term is
$-(\I\kb/\pi)\pvint_{-1}^{\infty} m(s)/(\xi-s)\,\D s$, taken over the chord and
the wake together. Over the chord alone it would not be bounded: $m(1) = \Gamma$
is not zero, and the end point $s = 1$ contributes a logarithm, $\log|1-\xi|$,
as $\xi \to 1$. The wake integral contributes the same logarithm with the
opposite sign, because $m$ is continuous across the trailing edge, $m(1^-) =
m(1^+) = \Gamma$, and the two cancel. Over $[-1,\infty)$ the function $m$ is
H\"older-continuous, vanishes at the leading edge and is continuous at $\xi =
1$, so the term is bounded on the whole chord, trailing edge included. The
Hankel terms acting on the wake are bounded for the same reason. Hence
$\mathcal{K}$, which includes the wake through the circulation, is a bounded
\emph{operator}, made up of the regularised $H_1^{(1)}$ acting on $\gamma$, the
$k_e^2H_0^{(1)}$ acting on $m$ and the Hilbert transform of $m$ over the chord
and the wake, and the dominant singular part of \eqref{eq:sie} is $k_e$-independent, which is why the
exponents of \S\ref{sec:edge} are the same at every Mach number. Setting
$k_e = 0$ removes both Hankel terms and leaves the incompressible equation of
\citet{baddoo2021}.

\section{The trailing-edge exponent in the basis}
\label{app:te}

The basis \eqref{eq:basis} carries the trailing-edge exponent $\alpha$ of
\eqref{eq:alpha} as a parameter, so that a porous trailing edge is represented
exactly rather than through a rigid-edge weight. Three points are worth
recording.

First, the wake mode $g_K = ((1+\xi)/2)^{1-\beta}$ carries no $(1-\xi)$ factor
and is therefore unchanged by $\alpha$; the tie $c_K = \I\kb\Gamma$ follows from
continuity of the wake and not from the edge exponent, so \eqref{eq:kutta} holds
for every $\alpha > 0$.

Second, the closed-form antiderivatives $G_n$ generalise with the Jacobi
parameter: the whole-chord integrals are
$2^{\alpha-\beta+1}B(1-\beta, 1+\alpha)$ with $B$ the beta function, and the
orthogonality $\int g_n = 0$ for $n \ge 2$ that the solver relies on is
preserved. These have been checked against direct quadrature to
$10^{-7}$--$10^{-9}$ for real and complex $\alpha$.

Third, the generalisation is exact at $\alpha = 1/2$: the implementation
carrying general $\alpha$ reproduces the rigid-trailing-edge implementation
bit for bit, with zero difference in lift, circulation, loading distribution and
every expansion coefficient, over eleven cases spanning rigid plates, porous
nose inserts with junctions, complex $\lambda$, $k_e$ up to $4.4$, $\kb$ up to
$101$, and heave forcing.

How much does $\alpha$ matter? Less than one would expect for integrated
quantities, and a great deal for local ones. On the profiles of
figure~\ref{fig:bdcode}, where $\alpha \approx 0.15$, carrying the correct
exponent changes the lift by at most $2\times10^{-4}$; on a uniformly permeable
plate at $\lambda = 3$, where $\alpha = 0.102$ against the rigid $1/2$, the lift
agrees to five figures. The loading itself does not: within $10^{-3}$ of the
trailing edge the two bases differ by a factor of order five, in the direction
\eqref{eq:alpha} requires. The reason integrated quantities are insensitive is
that the region in which the two exponents differ materially is that small.
Local quantities are a different matter, and the loading near a permeable
trailing edge is exactly what a trailing-edge scattering calculation needs; that
is why the exponent is carried as a parameter rather than fixed.

\section{Quadrature and the aliasing condition}
\label{app:quad}

Each influence integral is evaluated on Gauss--Legendre panels graded
geometrically towards the singular points $\xi = \pm1$ and $\xi = s_j$, with the
Cauchy singularity removed by the subtraction \eqref{eq:subtract}. Grading alone
is insufficient: the basis members oscillate as $\cos(N\vartheta)$, so a panel
wider than one half-period aliases, and the resulting error is systematic rather
than random. Requiring every panel to satisfy
$\Delta\vartheta \le \pi n_p/(4N)$ removes it, and the junction-invariance test
of table~\ref{tab:verify} falls from $7\,\%$ to $3.5\times10^{-10}$ when the
condition is imposed. The grading depth is capped so that panel edges never fall
below floating-point resolution.

\section{Steady closed forms}
\label{app:steady}

At $\omega = 0$ the dominant equation is
$\lambda\Pi_0(\xi) + \pi^{-1}\pvint\Pi_0(\tau)/(\xi-\tau)\,\D\tau = -\alpha_i$
on $[-1,1]$, with bounded loading at the trailing edge. Seeking
$\Pi_0 = A_0[(1-\xi)/(1+\xi)]^{\beta}$, which for a uniformly porous plate
carries $(1-\xi)^{+\beta}$ at the trailing edge (the bounded-loading condition
in that case, consistent with \eqref{eq:alpha} at $\lambda(1) = \lambda$), and
using the Tricomi identity \citep{tricomi1957}
\begin{equation}
\frac{1}{\pi}\pvint_{-1}^{1}
\Big(\frac{1-\tau}{1+\tau}\Big)^{\beta}\frac{\D\tau}{\tau-\xi}
= \cot(\pi\beta)\Big(\frac{1-\xi}{1+\xi}\Big)^{\beta} - \csc(\pi\beta)
\end{equation}
reduces the equation to the algebraic pair $\lambda = \cot(\pi\beta)$ and
$A_0 = -\alpha_i\sin(\pi\beta)$: the $\xi$-dependent terms cancel identically
only at that $\beta$, which is \eqref{eq:beta}, and the constant terms then fix
$A_0$. Integrating with
$\int_{-1}^{1}[(1-\xi)/(1+\xi)]^{\beta}\D\xi = 2\pi\beta/\sin(\pi\beta)$ gives
$\int\Pi_0\,\D\xi = -2\pi\beta\alpha_i$, so that $C_L = -2\int\Pi_0\,\D\xi =
4\pi\beta\alpha_i$ and hence \eqref{eq:CL}. The seepage dissipation is
$-\int v_s\Delta p\,\D x_1$, which is positive and by the same identity equals $\alpha_iC_L$, that is, $2\beta$ times the leading-edge suction $2\pi\alpha_i^2$ of the
corresponding rigid plate, not equal to it, because a permeable edge with
$\beta < 1/2$ develops no suction.

\paragraph{The insert law \eqref{eq:CLinsert}.} Let the leading-edge insert
occupy $-1 < \xi < s_j$ with a constant $\lambda$ and exponent $\beta$, the rest
of the chord being rigid. Writing $\Omega(z) =
(2\pi\I)^{-1}\int\Pi_0(\tau)/(\tau - z)\,\D\tau$, the Plemelj formulae turn the
steady equation into $(\lambda - \I)\Omega^+ - (\lambda + \I)\Omega^- =
-\alpha_i$ on the chord, that is $\Omega^+ = G\,\Omega^- - \alpha_i/(\lambda -
\I)$ with $G$ of \eqref{eq:G}. For a constant upwash this has the constant
particular solution $\Omega = -\I\alpha_i/2$, whatever $\lambda$ is, because
$(\lambda - \I) - (\lambda + \I) = -2\I$. The canonical function
\[
X(z) = (z+1)^{-\beta}\,(z - s_j)^{\beta - 1/2}\,(z-1)^{1/2}
\]
has $X^+/X^- = \E^{2\pi\I\beta} = G$ on the insert, $X^+/X^- = -1 = G$ on the
rigid part, no jump off the chord, the edge behaviour $(1+\xi)^{-\beta}$, $|\xi
- s_j|^{\beta - 1/2}$ and $(1-\xi)^{1/2}$ of \eqref{eq:beta}, \eqref{eq:delta}
and the Kutta condition, and $X \to 1$ at infinity since its exponents sum to
zero. Hence $\Omega(z) = -(\I\alpha_i/2)\,[1 - X(z)]$ solves the problem and
vanishes at infinity; it is the only such solution, since the difference of two
solutions divided by $X$ is entire and vanishes at infinity. As $z \to \infty$,
$\Omega \sim -(2\pi\I z)^{-1}\int\Pi_0\,\D\xi$ and $X = 1 + \kappa_1/z +
O(z^{-2})$ with $\kappa_1 = -\beta - (\beta - \tfrac12)s_j - \tfrac12 = -1 +
(\tfrac12 - \beta)(1 + s_j)$, which is linear in $s_j$. Therefore
\[
\int_{-1}^{1}\Pi_0\,\D\xi = \pi\alpha_i\kappa_1
= -\pi\alpha_i\Big[1 - \frac{\ell}{c}\,(1 - 2\beta)\Big],
\qquad \frac{\ell}{c} = \frac{1 + s_j}{2},
\]
which is \eqref{eq:CLinsert}: the rigid plate at $s_j = -1$, \eqref{eq:CL} at
$s_j = 1$, and exactly linear between. The argument extends to any number of
piecewise-constant segments, the lift remaining linear in each junction
position; it relies on the upwash being constant and does not carry over to a
cambered section.

\section{Inversion to the time domain}
\label{app:indicial}

What is inverted is the lift transfer function with its apparent-mass part
removed. For heave that part is an impulse at $\tau = 0$ and is not what
Wagner's function describes: writing the computed response as
$L(\sigma) \sim a - \I b\sigma$ at the top of the sweep, the fitted $b$ is
subtracted, $H(\sigma) = [L(\sigma) + \I b\sigma]/L(0)$. The fit returns
$b = 0.5006$ for a rigid plate, against the exact $1/2$, and $|b| \le 0.003$ for
every permeable case computed. A permeable plate has almost no apparent mass
(\S\ref{sec:theo}), so for those cases the subtraction is immaterial. At
$M = 0.5$ it returns $|b| < 2\times10^{-3}$ even for a rigid plate, whose
apparent mass compressibility has already removed. For the
gust no subtraction is needed, and the leading-edge phase reference of
\eqref{eq:forcing} is the one K\"ussner's function requires. At finite Mach
number both transfer functions are formed from the physical lift
\eqref{eq:lift}; the transformed integral of $\Pi$ would shift the gust
reference by the constant phase $Mk_e$ and misstate the heave response.

The inversion \eqref{eq:indicial} is then split at the top of the computed
range, $\Sigma = 25$:
\[
\phi(\tau) = \frac{2}{\pi}\left[\int_{\sigma_{\min}}^{\Sigma}
\frac{\operatorname{Re}H(\sigma)}{\sigma}\sin(\sigma\tau)\,\D\sigma
\;+\; c_\infty\Big(\frac{\pi}{2} - \mathrm{Si}(\Sigma\tau)\Big)\right],
\qquad c_\infty = \lim_{\sigma\to\Sigma}\operatorname{Re}H ,
\]
the tail being exact for a transfer function that has settled to a constant.
$H$ is computed on $102$ points, of which $48$ are geometrically spaced from
$\sigma_{\min} = 10^{-4}$ to $0.5$, $40$ are uniformly spaced up to $8$ and $14$
are geometrically spaced up to $25$. It is splined onto a
grid fine enough to carry at least twenty points per half-period of
$\sin(\sigma\tau)$ before integration; without that refinement the long-time
values are aliased. Three convergence checks were made: lowering
$\sigma_{\min}$ from $10^{-3}$ to $10^{-4}$ changes $\phi(40)$ by $0.02$ and
$\phi(\tau \le 10)$ by less than $0.006$; raising $\Sigma$ from $25$ to $50$,
for the rigid and $\lambda_0 = 3$ Wagner and K\"ussner analogues at $M = 0$ and
$0.5$, changes $\phi$ by less than $10^{-3}$ for $\tau \ge 1$ in every case and
by at most $0.014$ below $\tau = 1$, the largest change being in the rigid
Wagner analogue at $M = 0.5$, whose initial piston value is approached only as
$\Sigma$ grows; and the rigid Wagner function is reproduced to $0.007$ over
$\tau \le 40$, which is the end-to-end test of the scheme.

\section{An independent solution in physical variables}
\label{app:indep}

The check of \S\ref{sec:verify} solves the same physical problem by a route
that shares no variable, basis or closure with \S\ref{sec:numerics}. With
$b = U = \rho_0 = 1$ and $\E^{-\I\omega t}$, a pressure jump
$\Delta p(x_1) = p(x_1,0^-) - p(x_1,0^+)$ carried on the chord alone induces
the normal velocity
\[
v(x_1) = \frac{1}{2\pi}\int_{-\infty}^{\infty}\hat K(\zeta)\,
\widehat{\Delta p}(\zeta)\,\E^{\I\zeta x_1}\,\D\zeta ,
\qquad
\hat K(\zeta) = \frac{-\I\,\nu(\zeta)}{2(\sigma + \I0 - \zeta)},
\qquad
\nu = \big[\zeta^2 - (k_0 - M\zeta)^2\big]^{1/2},
\]
which follows from the convected wave equation for a potential odd in $x_2$.
The branch of $\nu$ has $\operatorname{Re}\nu \ge 0$ and is
$-\I|\zeta^2 - (k_0-M\zeta)^2|^{1/2}$ on the radiating interval, and the
$+\I0$, which is causality, makes the pole at $\zeta = \sigma$ contribute
$-\I\pi$ times its residue; on the real axis the rest is a principal value,
evaluated with a Gauss rule symmetric about the pole. No wake appears, because
the pressure is continuous across it, and the Kutta condition is imposed by the
basis. The trial functions are
$(1+x_1)^{-\beta}(1-x_1)^{\alpha}P_n^{(\alpha,-\beta)}(x_1)$ with the exponents
of \S\ref{sec:edge}, their Fourier transforms computed by Gauss--Jacobi
quadrature (for a rigid plate, $\alpha = \beta = 1/2$, they are Bessel
functions in closed form, and both versions were run); the test functions are
$\sin m\vartheta$, $x_1 = -\cos\vartheta$. The permeable condition adds
$(\lambda\beta_M/2)\Delta p$ to the normal velocity, which is the seepage of
\eqref{eq:vs} rewritten with \eqref{eq:lambda}; the sign differs from
\eqref{eq:vs} because $\Delta p$ is here lower minus upper.
The imposed velocity is the physical upwash, the lift is
$\int\Delta p\,\D x_1$, and nothing is transformed at either end. The
$\zeta$-integral is truncated at $|\zeta| = 2500$--$6000$; for a rigid plate
the integrand decays only as $\zeta^{-2}$ and the truncation limits the
accuracy to about $10^{-3}$ (the method reproduces the Sears and Theodorsen
functions at $M = 0$ to that level), while for a permeable plate the weaker
edge singularity makes it decay faster and the two solutions agree to
$10^{-5}$.

\bibliography{refs}

\end{document}